\pdfoutput=1

\documentclass[11pt,twoside,a4paper,cmspaper,final,collab]{cms-tdr}
\def\svnVersion{488075P}\def\cmsCernNoTag{CERN-EP-2018-250}\def\cmsCernDate{\today}\def\cmsMessage{Published in the European Physical Journal C as \href{http://dx.doi.org/10.1140/epjc/s10052-019-6562-5}{\doi{10.1140/epjc/s10052-019-6562-5.}}}

\begin{document}\cmsNoteHeader{SMP-17-012}

\hyphenation{had-ron-i-za-tion}
\hyphenation{cal-or-i-me-ter}
\hyphenation{de-vices}
\RCS$HeadURL: svn+ssh://svn.cern.ch/reps/tdr2/papers/SMP-17-012/trunk/SMP-17-012.tex $
\RCS$Id: SMP-17-012.tex 488025 2019-02-01 05:53:42Z hjheng $

\providecommand{\cmsTable}[1]{\resizebox{\textwidth}{!}{#1}}
\providecommand{\NA}{\ensuremath{\text{---}}}

\newlength\cmsFigWidth
\ifthenelse{\boolean{cms@external}}{\setlength\cmsFigWidth{0.85\columnwidth}}{\setlength\cmsFigWidth{0.4\textwidth}}
\ifthenelse{\boolean{cms@external}}{\providecommand{\cmsLeft}{upper\xspace}}{\providecommand{\cmsLeft}{left\xspace}}
\ifthenelse{\boolean{cms@external}}{\providecommand{\cmsRight}{lower\xspace}}{\providecommand{\cmsRight}{right\xspace}}

\cmsNoteHeader{SMP-17-012}

\title{Search for rare decays of $\cPZ$ and Higgs bosons to $\JPsi$ and a photon in proton-proton collisions at $\sqrt{s}$ = 13\TeV}

\date{\today}

\abstract{
A search is presented for decays of $\cPZ$ and Higgs bosons to a $\JPsi$ meson and a photon, with the subsequent decay of the $\JPsi$ to $\MM$. The analysis uses data from proton-proton collisions with an integrated luminosity of 35.9\fbinv at $\sqrt{s}=13\TeV$ collected with the CMS detector at the LHC. The observed limit on the $\cPZ\to\JPsi\gamma$ decay branching fraction, assuming that the $\JPsi$ meson is produced unpolarized, is $1.4\times10^{-6}$ at 95\% confidence level, which corresponds to a rate higher than expected in the standard model by a factor of 15. For extreme-polarization scenarios, the observed limit changes from -13.6 to +8.6\% with respect to the unpolarized scenario. The observed upper limit on the branching fraction for $\PH\to\JPsi\gamma$ where the $\JPsi$ meson is assumed to be transversely polarized is $7.6\times 10^{-4}$, a factor of 260 larger than the standard model prediction. The results for the Higgs boson are combined with previous data from proton-proton collisions at $\sqrt{s}=8\TeV$ to produce an observed upper limit on the branching fraction for $\PH\to\JPsi\gamma$ that is a factor of 220 larger than the standard model value.
}

\hypersetup{%
pdfauthor={CMS Collaboration},%
pdftitle={Search for rare decays of Z and Higgs bosons to J/Psi and a photon in proton-proton collisions at sqrt(s) = 13 TeV},%
pdfsubject={Rare decays},%
pdfkeywords={CMS, standard model physics, Z boson, Higgs boson, rare decays}}

\maketitle

\section{Introduction}

A new boson with a mass of 125\GeV was observed in data from the ATLAS and CMS experiments at the CERN LHC~\cite{Chatrchyan2013,Aad:2013xqa,201230,Chatrchyan:2013lba,CMS:2014ega,AtlasProperties,CMS:2015kwa}. All measurements of the properties of this boson are consistent with those of the Higgs boson ($\PH$) of the standard model (SM). However, the Yukawa couplings of the Higgs boson to the first- and second-generation quarks are currently only weakly constrained. Rare exclusive decays of the Higgs boson to mesons in association with a photon can be used to explore such couplings. For example, the $\PH\to\JPsi\gamma$ decay can probe the Higgs boson coupling to the charm quark~\cite{HiggsBosonDecaysToQuarkonia}. The corresponding decay, $\cPZ\to\JPsi\gamma$, can be used as an experimental benchmark in the search for $\PH\to\JPsi\gamma$~\cite{GUBERINA1980317,PhysRevD.92.014007}, and in checking approaches to factorization in quantum chromodynamics (QCD) used to estimate branching fractions ($\mathcal{B}$) in radiative decays of electroweak bosons~\cite{Grossmann:2015lea}.

\begin{figure*}[!ht]
\centering
\includegraphics[width=0.24\textwidth]{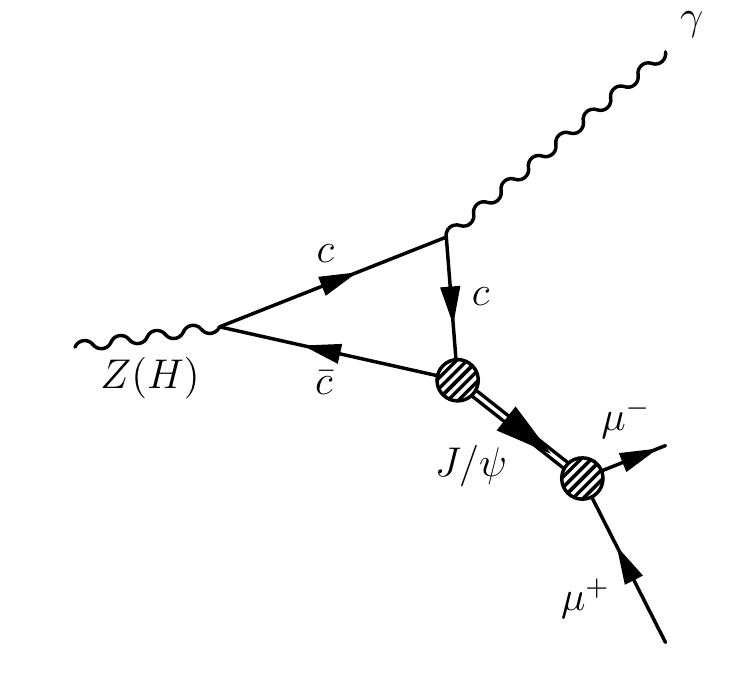}~
\includegraphics[width=0.24\textwidth]{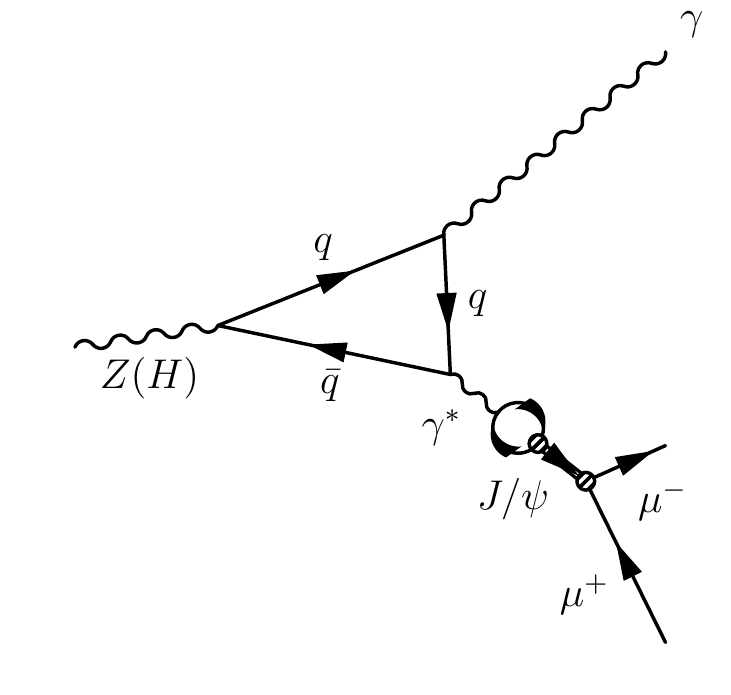}~
\includegraphics[width=0.24\textwidth]{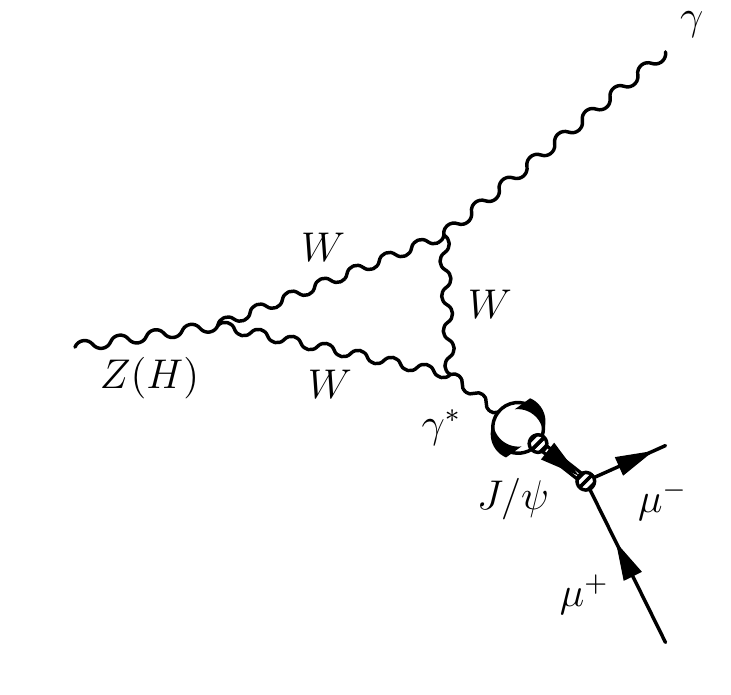}~
\includegraphics[width=0.24\textwidth]{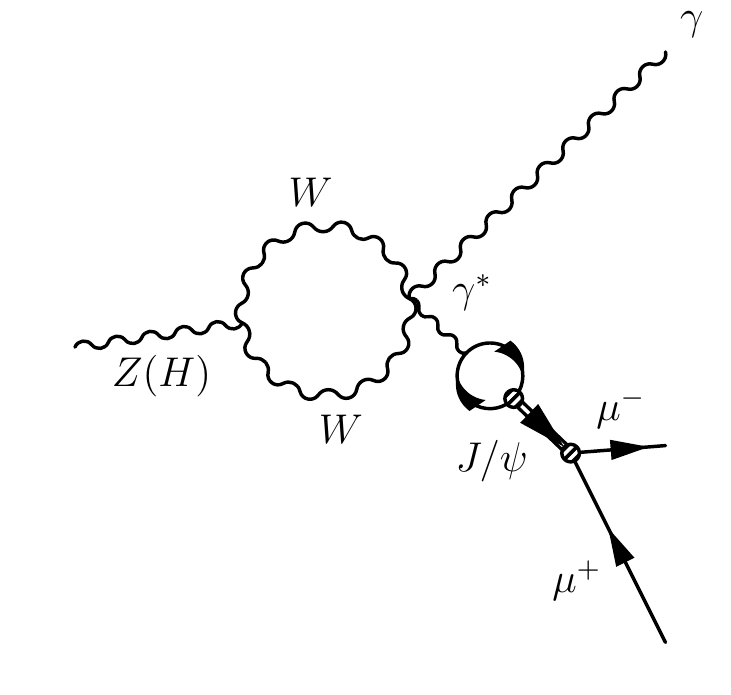}
\caption{Lowest order Feynman diagrams for the $\cPZ$ (or $\PH$)$\to\JPsi\gamma$ decay. The left-most diagram shows the direct and the remaining diagrams the indirect processes.}
\label{fig:ZJpsiG_diag}
\end{figure*}

Both $\cPZ$ and Higgs boson decays receive contributions from direct and indirect processes.  In the direct process, $\cPZ$ and Higgs bosons couple to charm quarks, and charm quarks then hadronize to form $\JPsi$ mesons. In the indirect process, the $\cPZ$ and Higgs bosons decay through quark or $\PW$ boson loops to $\gamma\gamma^{*}$, and the $\gamma^{*}$ then converts to a $\ccbar$ resonant state. The lowest order Feynman diagrams for these decay modes are shown in Fig.~\ref{fig:ZJpsiG_diag}. The latest SM calculations of the branching fractions of both decays, taking into account the interference between direct and indirect processes, are~\cite{PhysRevD.97.016009, PhysRevD.96.116014}:
\begin{linenomath}
\begin{equation}
\mathcal{B}_{\text{SM}}(\cPZ\to\JPsi\gamma)=(9.0^{+1.5}_{-1.4})\ten{-8},
\end{equation}
\begin{equation}
\mathcal{B}_{\text{SM}}(\PH\to\JPsi\gamma)=(3.0^{+0.2}_{-0.2})\ten{-6}.
\end{equation}
\end{linenomath}
Modified $\PH \ccbar$ couplings can arise in certain extensions of the SM~\cite{Delaunay:2013pja}. For example, within the context of effective field theory, the $\PH\ccbar$ coupling may be modified in the presence of a dimension-six operator, leading to an enhancement of coupling relative to the SM at the cutoff scale $\Lambda$  that can be as small as 30\TeV. This provides no other signature of new physics at the LHC. In the two Higgs doublet model with minimal flavor violation~\cite{Trott:2010iz,Jung:2010ik}, the $\PH \ccbar$ coupling can be significantly enhanced by breaking flavor symmetry, while other couplings are not severely affected. The composite pseudo-Nambu-Goldstone boson model~\cite{Giudice:2007fh} parametrizes the coupling by the degree of compositeness and compositeness scale. The coupling can be constrained through a direct experimental search for the composite particles associated with the charm quark~\cite{Delaunay:2013pwa}.

Deviations from SM predictions for the couplings can affect the interference terms and result in changes to the branching fractions. For example, the shift in the branching fraction for $\PH\to\JPsi\gamma$ can be more than 100\% if the $\PH \ccbar$ coupling deviates from its SM value by more than a factor of 2~\cite{HiggsBosonDecaysToQuarkonia}. Since this Higgs boson decay is sensitive to the $\PH\ccbar$ coupling, a measurement of the branching fraction can verify whether the Higgs boson couples to second-generation quarks with the strength predicted by the SM.

The ATLAS experiment has searched for the decay $\cPZ \to\JPsi\gamma$ in proton-proton ($\Pp\Pp$) collisions collected at $\sqrt{s}=8\TeV$~\cite{Aad:2015sda}. The respective observed and expected upper limits at 95\% confidence level (\CL) on the branching fraction were reported to be 2.6 and $2.0^{+1.0}_{-0.6}\ten{-6}$, where the subscript and superscript reflect the range in the 68\% central-quantiles of upper limits assuming a background-only hypothesis. Searches for the $\PH\to\JPsi\gamma$ decay were performed by ATLAS and CMS in $\Pp\Pp$ collisions collected at $\sqrt{s}=8\TeV$~\cite{Aad:2015sda,Run1Paper_Dalitz}. The respective observed and expected upper limits in the branching fractions were 1.5 and $1.2^{+0.6}_{-0.3}\ten{-3}$ from ATLAS, and 1.5 and $1.6^{+0.8}_{-0.8}\ten{-3}$ from CMS. The ATLAS experiment performed similar searches for both the $\cPZ$ and Higgs boson decays in $\Pp\Pp$ collisions collected at $\sqrt{s}=13\TeV$. The respective observed and expected upper limits on the branching fractions were 2.3 and $1.1^{+0.5}_{-0.3}\ten{-6}$ for the $\cPZ$ boson decay, and 3.5 and $3.0^{+1.4}_{-0.8}\ten{-4}$ for the Higgs boson decay~\cite{Aaboud:2018txb}. The ATLAS experiment also searched for the $\PH\to\ccbar$ decay in $\Pp\Pp \to \cPZ\PH$ production in data collected at $\sqrt{s}=13\TeV$~\cite{Aaboud:2018fhh}, and reported observed and expected limits on the ratio $\sigma(\Pp\Pp\to\cPZ\PH)\times\mathcal{B}(\PH\to\ccbar)$ relative to the SM prediction of 110 and $150^{+80}_{-40}$ respectively, where $\sigma(\Pp\Pp\to\cPZ\PH)\times\mathcal{B}(\PH\to\ccbar)$ is the upper limit for the cross section.

The results presented in this paper are based on $\Pp\Pp$ collisions at $\sqrt{s}=13\TeV$ recorded with the CMS detector, corresponding to an integrated luminosity of 35.9\fbinv.

\section{The CMS detector}
A detailed description of the CMS detector, together with a definition of the coordinate system used and the relevant kinematic variables, can be found in Ref.~\cite{CMS-Jinst}.  The central feature of the CMS apparatus is a superconducting solenoid, 13\unit{m} in length and 6\unit{m} in internal diameter, providing an axial magnetic field of 3.8\unit{T}. Within the solenoid volume are a silicon pixel and strip tracker, a lead tungstate crystal electromagnetic calorimeter (ECAL), and a brass and scintillator hadron calorimeter (HCAL), each composed of a barrel and two endcap sections. Forward calorimeters extend the pseudorapidity ($\eta$) coverage provided by the barrel and endcap detectors. Muons are detected in gas-ionization chambers embedded in the steel flux-return yoke outside the solenoid.

The silicon tracker measures charged particles within the range $\abs{\eta} < 2.5$. It consists of 1440 silicon pixel and 15\,148 silicon strip detector modules. For non-isolated particles with transverse momentum, \pt, between 1 and 10\GeV and $\abs{\eta} < 1.4$, the track resolutions are typically 1.5\% in \pt and 25--90 (45--150)\unit{$\mu$m} in the transverse (longitudinal) direction~\cite{TRK-11-001}.

The ECAL consists of 75\,848 crystals, which provide coverage in $\abs{\eta} < 1.479$ in the barrel region (EB) and $1.479< \abs{\eta} < 3.000$ in the two endcap regions (EE). The preshower detectors, each consisting of two planes of silicon sensors interleaved with a total of $3X_{0}$ of lead are located in front of the EE~\cite{Khachatryan:2015hwa,CMS:EGM-14-001}. In the barrel section of the ECAL, an energy resolution of about 1\% is achieved for unconverted or late-converting photons in the tens of \GeV energy range. The remaining barrel photons have a resolution of about 1.3\% up to $\abs{\eta} = 1$, rising to about 2.5\% at $\abs{\eta} = 1.4$. In the endcaps, the resolution of unconverted or late-converting photons is about 2.5\%, while the remaining endcap photons have a resolution between 3 and 4\%~\cite{CMS:EGM-14-001}.

Muons are measured in the range $\abs{\eta} < 2.4$, with detection planes made using three technologies: drift tubes, cathode strip chambers, and resistive plate chambers. Matching muons to tracks measured in the silicon tracker results in a relative \pt resolution, for muons with \pt up to 100\GeV, of 1\% in the barrel and 3\% in the endcaps. The \pt resolution in the barrel is better than 7\% for muons with \pt up to 1\TeV~\cite{Sirunyan:2018fpa}.

A two-tier trigger system selects collision events of interest. The first level (L1) of the CMS trigger system~\cite{Khachatryan:2016bia}, composed of custom hardware processors, uses information from the calorimeters and muon detectors to select the most interesting events in a fixed time interval of less than 4\unit{$\mu$s}. The high-level trigger processor farm further decreases the event rate from around 100\unit{kHz} to less than 1\unit{kHz}, before data storage.

\section{Data and simulated samples}
\label{sec:sample}
The L1 trigger requires the presence of a muon with \pt greater than 5\GeV and an isolated electromagnetic object with \pt greater than 18\GeV. The HLT algorithm requires the presence of a muon and a photon with \pt exceeding 17 and 30\GeV, respectively. No isolation requirement is imposed on the muons because of the small angular separation expected between the muons in signal events. No further isolation constraint is required for the photon. The trigger efficiency for events satisfying the selection used in the analysis is determined using a high-purity ($\sim$97\%) $\cPZ\to\mu\mu\gamma$ control sample; it is measured to be $82\pm 0.7\%$ in data and $83\pm 0.4\%$ in simulated events.

Simulated samples of the $\cPZ$ and Higgs boson decays are used to estimate the expected signal yields and model the kinematic distributions of signal events. The $\cPZ\to\JPsi\gamma\to\mu\mu\gamma$ sample, with $m_{\cPZ}=91.2\GeV$~\cite{PDG2018}, is produced with the \PYTHIA 8.226 Monte Carlo (MC) event generator~\cite{SJOSTRAND2008852,Sjostrand:2014zea}, with hadronization and fragmentation using underlying event tune CUETP8M1~\cite{Khachatryan:2015pea}. The parton distribution function (PDF) set used is NNPDF3.0~\cite{Ball:2014uwa}. The SM $\cPZ$ boson production cross section includes the next-to-next-to-leading order (NNLO) QCD contributions, and the next-to-leading order (NLO) electroweak corrections from \FEWZ3.1~\cite{Li:2012wna} calculated using the NLO PDF set NNPDF3.0. The $\cPZ$ boson $\pt$ is reweighted to match the NLO calculation~\cite{Alioli:2008tz,Nason:2009ai,Alwall:2014hca}.

The $\PH\to\JPsi\gamma\to\mu\mu\gamma$ sample with $m_{\PH}=125\GeV$ is produced with the \POWHEG v2.0 MC event generator~\cite{Alioli:2008tz,Nason:2009ai} and includes gluon-gluon fusion ($\Pg\Pg$F), vector boson fusion (VBF), associated vector boson production (V$\PH$), and associated top quark pair production ($\ttbar\PH$). The generator is interfaced with \PYTHIA 8.212~\cite{SJOSTRAND2008852,Sjostrand:2014zea} for hadronization and fragmentation with tune CUETP8M1. The PDF set used is NNPDF3.0. The SM Higgs boson cross section is taken from the LHC Higgs cross section working group recommendations~\cite{LHC-YR4}.

In the SM, the $\JPsi$ meson from the Higgs boson decay must be fully transversely polarized in helicity frame ($\lambda_{\theta}=+1$, as described in Ref.~\cite{PhysRevD.83.031503}), because the Higgs boson has spin 0, and the photon is transversely polarized. Since the polarization of the $\JPsi$ meson is not correctly simulated in the signal samples, a reweighting factor is applied to each event to emulate the effect of polarization. The reweighting procedure results in a decrease of the signal acceptance by 7.0\%. For the $\cPZ$ boson decay, the helicity of the $\JPsi$ meson depends on that of the $\cPZ$ boson, which can have multiple helicity states. The results from the $\cPZ$ boson polarization measurement~\cite{Khachatryan:2015paa,Aad:2016izn} are not used to constrain the helicity of the $\JPsi$ meson in this analysis. The nominal results are obtained using a signal acceptance calculated for the unpolarized case. Assuming that the $\JPsi$ is produced with full transverse or longitudinal polarization ($\lambda_{\theta}$ = +1 or -1) changes the acceptance by -7.8\% or +15.6\%, respectively.

\begin{figure}[!htb]
\centering
\includegraphics[width=0.25\textwidth]{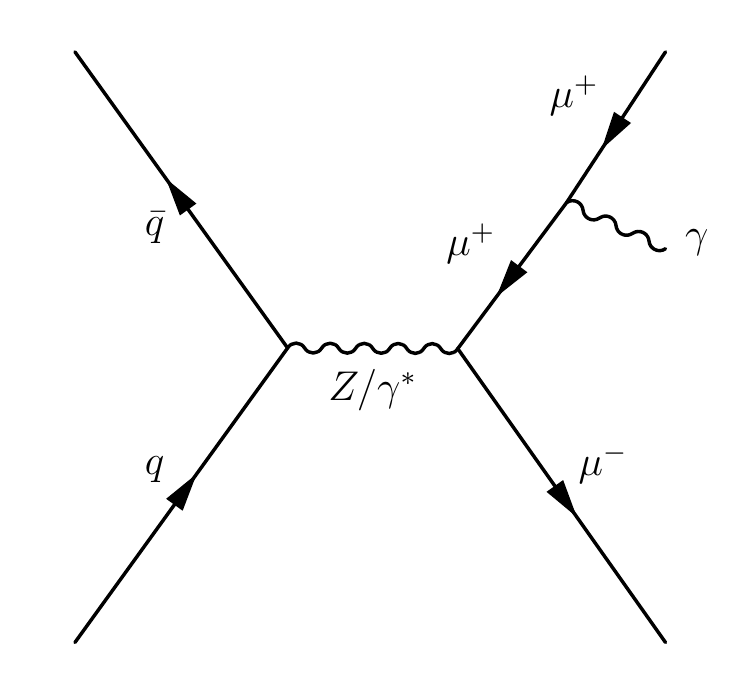}
\includegraphics[width=0.25\textwidth]{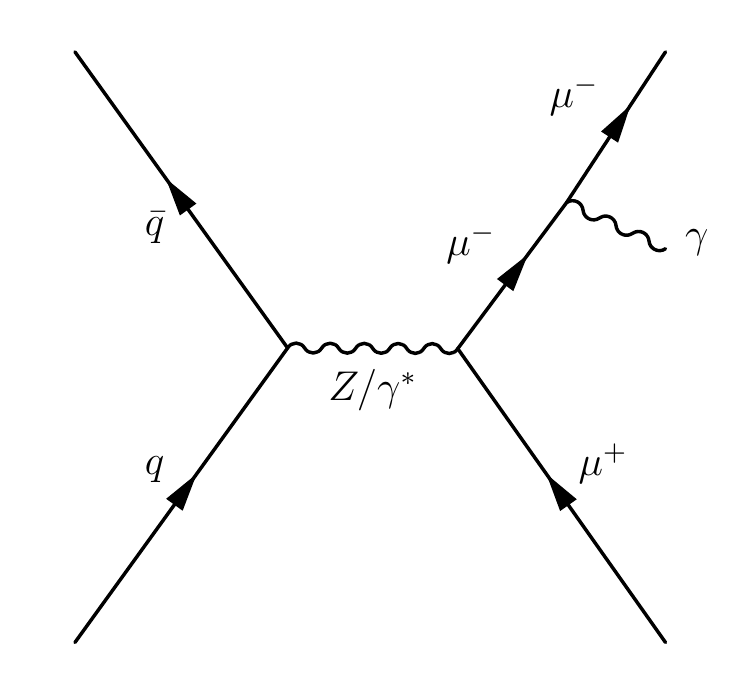}
\caption{The lowest order Feynman diagrams for the Drell-Yan process in $\Pp\Pp\to\cPZ\to\mu\mu\gamma$. The background exhibits a peak in $m_{\mu\mu\gamma}$ at the $\cPZ$ boson mass.}
\label{fig:FeynmanDiagrams_Zmmg}
\end{figure}

The Drell-Yan process, $\Pp\Pp\to\cPZ\to\mu\mu\gamma$, produces the same final state as the signal.
This process exhibits a peak at the $\cPZ$ boson mass, $m_{\cPZ}$, in the three-body invariant mass, $m_{\mu\mu\gamma}$,  as do the signal events, and it is therefore referred to as a resonant background. This background is included when deriving the upper limit on the branching fraction for $\cPZ \to\JPsi\gamma$. The lowest order Feynman diagrams for the $\Pp\Pp\to\cPZ\to\mu\mu\gamma$ process are shown in Fig.~\ref{fig:FeynmanDiagrams_Zmmg}. The \MGvATNLO 2.6.0 matrix element generator~\cite{Alwall:2014hca} is used to generate a sample of these resonant background events at leading order with the NNPDF3.0 PDF set, interfaced with \PYTHIA 8.226 for parton showering and hadronization with tune CUETP8M1. The photons in these events are all produced in final-state radiation from the $\cPZ\to\mu\mu$ decay, and therefore the $m_{\mu\mu\gamma}$ distribution peaks at the $\cPZ$ boson mass without a continuum contribution.

\begin{figure}[!htb]
\centering
\includegraphics[width=0.25\textwidth]{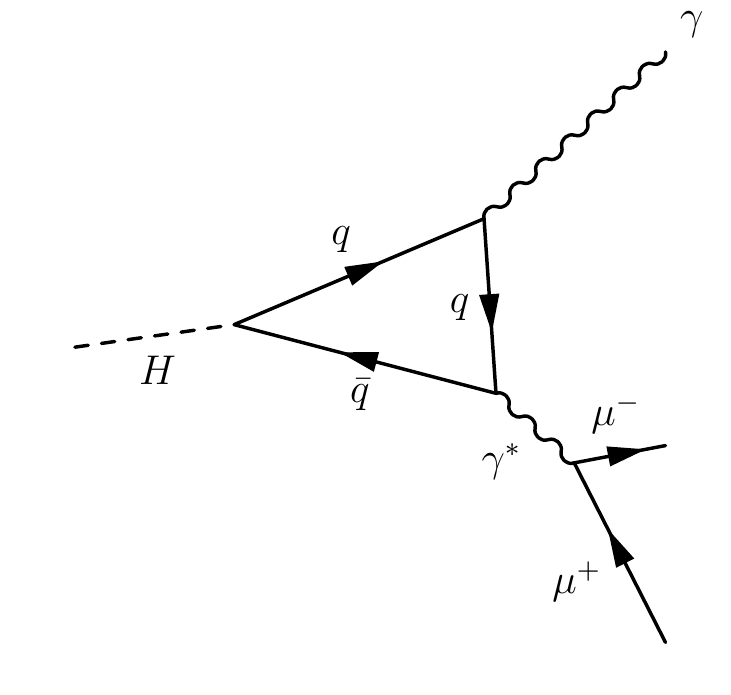} 
\includegraphics[width=0.25\textwidth]{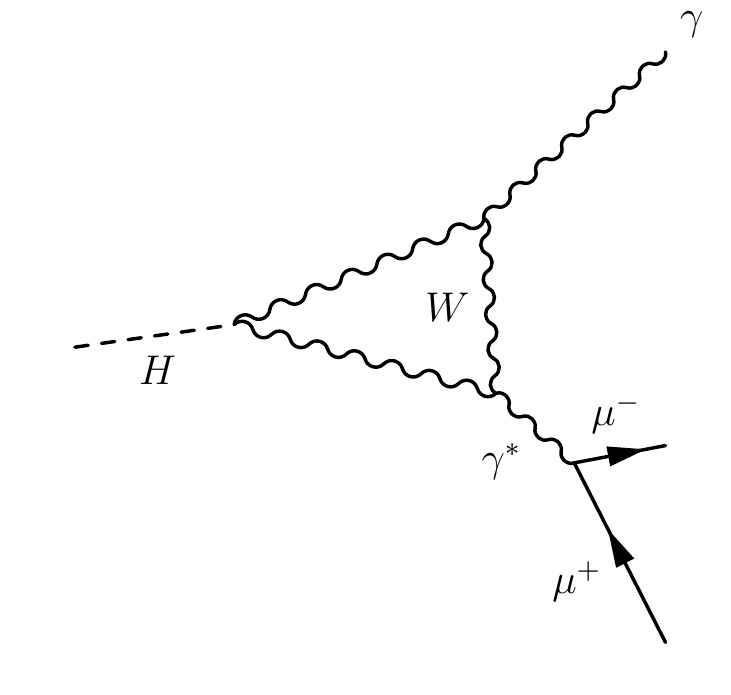} 
\includegraphics[width=0.25\textwidth]{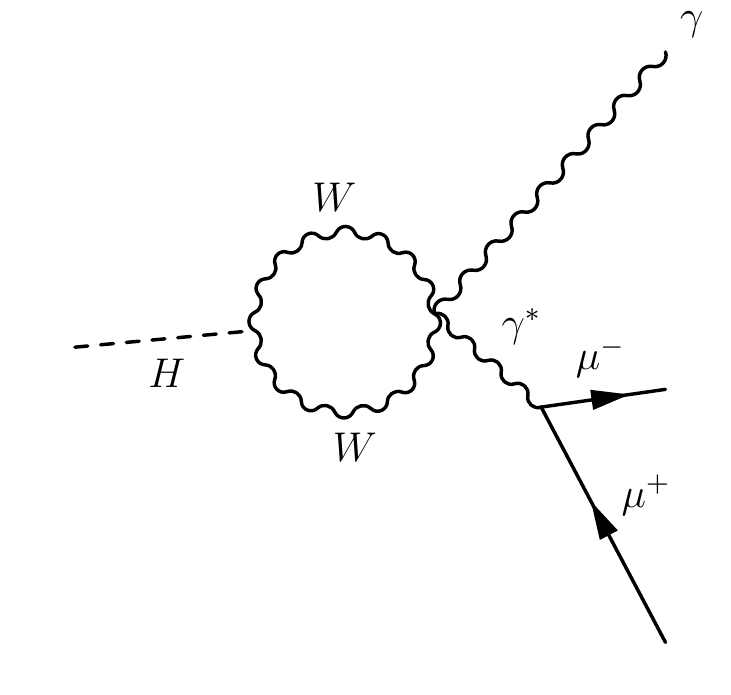}
\caption{The lowest order Feynman diagrams for the Higgs boson Dalitz decay of $\PH\to\gamma^{*}\gamma\to\mu\mu\gamma$. The background exhibits a peak in $m_{\mu\mu\gamma}$ at the Higgs boson mass.}
\label{fig:FeynmanDiagrams_Dalitz}
\end{figure}

Similarly, the Higgs boson Dalitz decay~\cite{Abba96}, $\PH\to\gamma^{*}\gamma\to\mu\mu\gamma$, is a resonant background to $\PH\to\JPsi\gamma$ decay. The lowest order Feynman diagrams for the $\PH\to\gamma^*\gamma$ process are shown in Fig.~\ref{fig:FeynmanDiagrams_Dalitz}. Samples of the Higgs boson Dalitz decays, produced via $\Pg\Pg$F, VBF, V$\PH$ modes for $m_{\PH}=125\GeV$, are simulated at NLO using the \MGvATNLO generator interfaced with \PYTHIA 8.212 for parton showering and hadronization. The $\ttbar\PH$ contribution is accounted for by scaling the VBF signal to the $\ttbar\PH$ production cross section. The branching fraction for $\PH\to\gamma^{*}\gamma$ is obtained from the \MCFM 7.0.1 program~\cite{MCFM7}. The other source of resonant background is the decay of a Higgs boson into two muons with a photon radiated from one of the muons. After the event selection, described in Section~\ref{sec:event}, the contribution of this background is negligible.

There are also background processes that do not give resonant peaks in the three-body invariant mass spectrum. These are referred to as nonresonant backgrounds. These processes include: (1) inclusive quarkonium production associated with either jets or photons where energetic jets can be misidentified as a photon ($\Pp\Pp\to \JPsi+\text{jets}/\gamma$), (2) the Drell-Yan process with associated jets ($\Pp\Pp\to\cPZ/\gamma^{*}+\text{jets}$), and (3) associated photons plus jets production ($\Pp\Pp\to\gamma+\text{jets}$). These nonresonant backgrounds, which are discussed in Section~\ref{sec:model}, are modeled using fits to the $m_{\mu\mu\gamma}$ distributions in data.

All generated events are processed through a detailed simulation of the CMS detector based on \GEANTfour~\cite{GEANT4}. Simultaneous $\Pp\Pp$ interactions that overlap the event of interest (pileup) are included in the simulated samples. The distribution of the number of additional pileup interactions per event in the simulation corresponds to that observed in the 13\TeV data collected in 2016.

\section{Event reconstruction and selection}
\label{sec:event}
The global event reconstruction (also called particle-flow event reconstruction~\cite{CMS-PRF-14-001}) reconstructs and identifies each individual particle in an event with an optimized combination of all subdetector information. In this process, the identification of the particle type (photon, electron, muon, charged hadron or neutral hadron) plays an important role in the determination of the particle direction and energy. Photons (\eg, coming from \Pgpz\ decays or from electron bremsstrahlung) are identified as ECAL energy clusters not linked to the extrapolation to the ECAL of any charged particle trajectory. Electrons are identified as a primary charged particle track with one or more ECAL energy clusters consistent with the extrapolation of this track to the ECAL or with bremsstrahlung photons emitted as the electron passes through the tracker material. Muons (\eg, from \cPqb-hadron semileptonic decays) are identified as a track in the central tracker consistent with either a track or several hits in the muon system, and associated with calorimeter deposits compatible with the muon hypothesis.  Charged hadrons are identified as charged particle tracks that are not identified as electrons or muons. Finally, neutral hadrons are identified as either HCAL energy clusters not linked to any charged hadron trajectory or ECAL and HCAL energy excesses with respect to any expected charged hadron energy deposit.

The high instantaneous luminosity of the LHC results in multiple $\Pp\Pp$ interactions per bunch crossing. The reconstructed vertex with the largest value of summed physics-object $\pt^2$ is the primary $\Pp\Pp$ interaction vertex. The physics objects are the jets, clustered using the anti-$\kt$ jet finding algorithm~\cite{Cacciari:2008gp,Cacciari:2011ma} with the tracks assigned to the vertex as inputs, and the associated missing \pt, taken as the negative vector \pt sum of those jets.

Photon and electron candidates are reconstructed by summing and clustering the energy deposits in the ECAL crystals. Groups of these clusters, called superclusters, are combined to recover the bremsstrahlung energy of electrons and converted photons passing through the tracker.  In the endcaps, preshower energy is added in the region covered by the preshower ($1.65<\abs{\eta} <2.60$). The clustering algorithms result in an almost complete recovery of the energy of photons.

A multivariate discriminant is used to identify photon candidates. The inputs to the discriminant are the isolation variables, the ratio of hadronic energy in the HCAL towers behind the superclusters to the electromagnetic energy in the superclusters, and the transverse width of the electromagnetic shower.  A conversion-safe electron veto~\cite{CMS:EGM-14-001}, which requires no charged-particle track with a hit in the inner layer of the pixel detector pointing to the photon cluster in the ECAL, is applied to avoid misidentifying an electron as a converted photon. Photons are required to be reconstructed within the region $\abs{\eta} < 2.5$, although those in the ECAL transition region $1.44<\abs{\eta}<1.57$ are excluded from the analysis. The efficiency of the photon identification procedure is measured with $\cPZ\to\re\re$ events using ``tag-and-probe'' techniques~\cite{cite:tagandprobe}, and is between 84--91 (77--94)\%, depending on the transverse energy $\et$, in the barrel (endcap). The electron veto efficiencies are measured with $\cPZ\to\mu\mu\gamma$ events, where the photon is produced by final-state radiation, and found to be 98 (94)\% in the barrel (endcap).

Muons are reconstructed by combining information from the silicon tracker and the muon system~\cite{Chatrchyan:2012xi}. The matching between the inner and outer tracks proceeds either outside-in, starting from a track in the muon system, or inside-out, starting from a track in the silicon tracker. In the latter case, tracks that match track segments in only one or two planes of the muon system are also included in the analysis to ensure that very low-\pt muons that may not have sufficient energy to penetrate the entire muon system are retained. Muons reconstructed only in the muon system are not retained for the analysis. In order to avoid reconstructing a single muon as multiple muons, whenever two muons share more than half of their segments, the one with lower reconstruction quality is removed. The compatibility with a minimum ionizing particle signature expected in the calorimeters is taken into account~\cite{Sirunyan2017}. Muons with $\pt>4\GeV$ and $\abs{\eta} <2.4$ are accepted.

To suppress muons originating from in-flight decays of hadrons, the impact parameter of each muon track, defined as its distance of closest approach to the primary event vertex position, is required to be less than 0.5 (1.0)\unit{cm} in the transverse (longitudinal) plane. In addition, the three-dimensional impact parameter is required to be less than four times its uncertainty. A cone of size $\Delta R = \sqrt{\smash[b]{(\Delta\phi)^2 +
(\Delta \eta)^2}} = 0.3$ is constructed around the momentum direction of each muon candidate, where $\phi$ is the azimuthal angle in radians.  The relative isolation variable for the muons is defined by summing the \pt of all photons, charged hadrons, and neutral hadrons within this cone, correcting for additional underlying event activity due to pileup events~\cite{Chatrchyan:2012vp}, and then dividing by the muon \pt:
\begin{linenomath}
\ifthenelse{\boolean{cms@external}}{
\begin{equation}\begin{split}
\label{eqn:pfiso}
\mathcal{I}^{\mu} \equiv& \Big( \sum \PT^\text{charged} \\ &+
                                 \max\big[ 0, \sum \PT^\text{neutral}
                                 + \sum \PT^{\Pgg}
                                 - \PT^\mathrm{PU}(\mu) \big] \Big)
                                 / \PT^{\mu},
\end{split}
\end{equation}
}{
\begin{equation}
\label{eqn:pfiso}
\mathcal{I}^{\mu} \equiv \Big( \sum \PT^\text{charged} +
                                 \max\big[ 0, \sum \PT^\text{neutral}
                                 +
                                  \sum \PT^{\Pgg}
                                 - \PT^\mathrm{PU}(\mu) \big] \Big)
                                 / \PT^{\mu},
\end{equation}
}
\end{linenomath}
where $\pt^{\text{PU}}(\mu)\equiv 0.5\sum_{i} \pt^{\text{PU},i}$, and $i$ runs over the momenta of the charged-hadron particle-flow candidates not originating from the primary vertex. The $\sum \PT^\text{charged}$ is the scalar \pt sum of charged hadrons originating from the primary event vertex. The $\sum \PT^\text{neutral}$ and $\sum \PT^{\Pgg}$ are the scalar  \pt sums of neutral hadrons and photons, respectively. The requirement $\mathcal{I}^{\mu} < 0.35$ is imposed on the leading muon to reject muons from electroweak decays of hadrons within jets or any jets that punch through the calorimeters mimicking a muon signature. The angular separation $\Delta R$ between the two muons is small because of their low invariant mass, $m_{\mu\mu}$, and the high \pt of the $\JPsi$ meson from the decay of the $\cPZ$ or Higgs boson. Therefore, no isolation requirement is applied to the subleading muons since they are within the isolation cone of the leading muon in most events. The momentum of the subleading muon is excluded from the isolation calculation. The efficiency of identification is measured in $\cPZ\to\mu\mu$ and $\JPsi\to\mu\mu$ events using the tag-and-probe method, and is 94--98 (92--97)\% in the barrel (endcap), depending on muon \pt and $\eta$. The isolation efficiency, which is $\pt$ dependent, is measured to be 90--100 (92--100)\% in the barrel (endcap), and is consistent with the measurement from $\cPZ\to\mu\mu$ events.

Signal candidates are selected by applying additional selection criteria to events containing at least two muons and one photon. The two muons must have opposite charges and $\pt>20\ (4)\GeV$ for the leading (subleading) muon. The \pt requirement for the leading muon is driven by the trigger threshold. The requirement that the photon has $\et >33\GeV$ is also driven by the trigger threshold. The angular separation of each muon from the photon is required to satisfy $\Delta R>1$ in order to suppress Drell-Yan background events with final-state radiation. To ensure that the dimuon $\JPsi$ candidate is well-separated from the photon, events are required to have $\Delta R(\mu\mu,\gamma) > 2$ and $\abs{\Delta\phi(\mu\mu,\gamma)}>1.5$. Both the photon and dimuon momenta must satisfy $\pt/m_{\mu\mu\gamma}>0.38\ (0.28)$ for the $\cPZ$ ($\PH$) boson decay. This constraint helps to reject the $\gamma^*+$jet and $\gamma+$jet backgrounds, with minimal effect on the signal efficiency and $m_{\mu\mu\gamma}$ spectrum. Events in which the mass of the two muons is consistent with the mass of the $\JPsi$ meson~\cite{PDG2018}, $3.0<m_{\mu\mu}<3.2\GeV$, are retained. In addition, only events with a three-body invariant mass in the range of $70\ (100) < m_{\mu\mu\gamma} < 120\ (150)\GeV$ are considered in the $\cPZ\ (\PH)$ boson search.

The simulated events are reconstructed using the same algorithms as the data, but the simulation does not reproduce the data perfectly. The differences in efficiencies between data and simulation for trigger, offline object reconstruction, identification, and isolation are corrected by reweighting the simulated events with data-to-simulation correction factors. The scale correction factors are observed to deviate from 1 by less than 2.5\%. The energy and momentum resolutions for muons and photons in simulated events are also corrected to match those in $\cPZ\to\mu\mu/\re\re$ events in data.

In the $\cPZ\to\JPsi\gamma$ search, selected events are classified into mutually exclusive categories in order to enhance the sensitivity of the search. The categorization is based on the $\eta$ and $\RNINE$ variables of the photon, where $\RNINE$ is defined as the energy sum of 3$\times$3 ECAL crystals centered on the most energetic crystal in the supercluster associated with the photon, divided by the energy of the supercluster~\cite{CMS:EGM-14-001}. Photons that do not convert to an $\EE$ pair in the detector tend to have high values of $\RNINE$ and a threshold of 0.94 is used to classify reconstructed photons with high $\RNINE$ (thus with a better resolution) and low $\RNINE$ (worse resolution). The three categories are: (1) photon in the barrel region with a high $\RNINE$ value (referred to as EB high $\RNINE$); (2) photon in the barrel region with low $\RNINE$ value (referred to as EB low $\RNINE$); and (3) photon in the endcap region (referred to as EE). The EE category is not divided into high/low $\RNINE$ because there are only a few events in this category. Events in the $\PH\to\JPsi\gamma$ search are not divided into categories since the sample size is limited and the sensitivity is still far from the SM prediction, and therefore event categorization does not result in a significant improvement in the expected limit.

Table~\ref{tab:Cutflow} shows the numbers of observed events in data, the expected yields from the $\cPZ\ (\PH)\to\JPsi\gamma$ signals, the expected nonresonant backgrounds with uncertainties estimated from the fits (described in Section~\ref{sec:model}), and the expected resonant background contributions in the range of $81\ (120) < m_{\mu\mu\gamma} < 101\ (130)\GeV$ for the $\cPZ\ (\PH)$ boson search. The values for the signal yields quoted for the $\cPZ$ boson decay assume that the $\JPsi$ meson is unpolarized and those for the Higgs boson decay assume transverse polarization for the $\JPsi$ meson. In the $\cPZ$ and Higgs boson channels, the numbers of events coming from the resonant backgrounds are large compared with those expected for the signal in the SM. However, the resonant backgrounds are small compared to the nonresonant backgrounds and therefore their effect on the final result is minimal.

The overall signal efficiency, including kinematic acceptance, trigger, object reconstruction, identification, and isolation efficiencies for the $\JPsi\gamma\to\mu\mu\gamma$ final state, is approximately 14 (22)\% for the $\cPZ$ ($\PH$) boson signal, respectively. The total signal efficiency for the $\cPZ$ boson decay is 13\% if the $\JPsi$ meson is fully transversely polarized and 16\% if it is fully longitudinally polarized. The difference between the efficiency for the $\cPZ$ boson and that for the Higgs boson arises from the differences in the $\pt$ spectra for the muons and the photon in the two cases. These differences are due to the difference between the $\cPZ$ boson and Higgs boson masses.

\begin{figure}[!ht]
	\centering
    \includegraphics[width=0.45\textwidth]{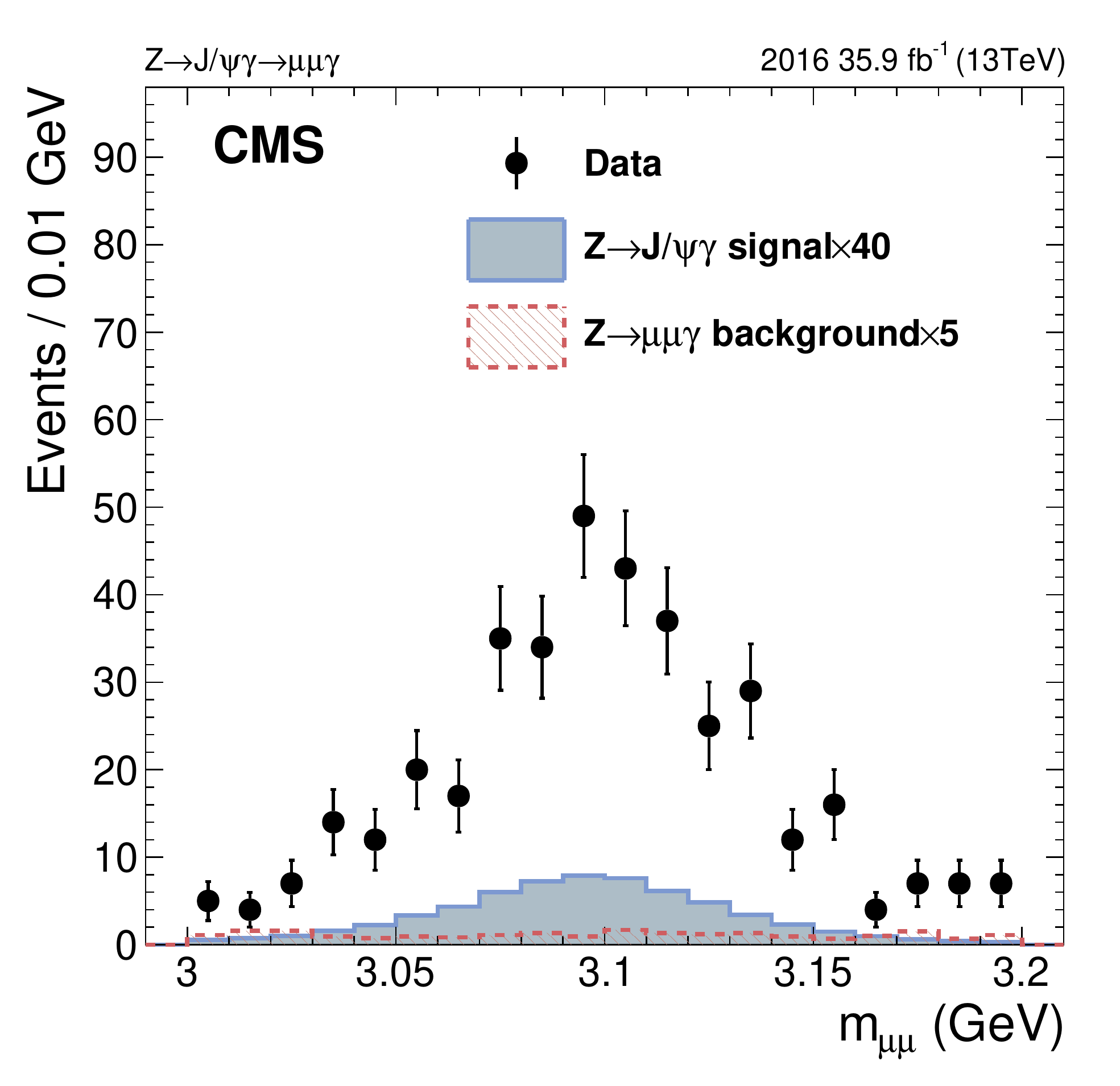}
    \includegraphics[width=0.45\textwidth]{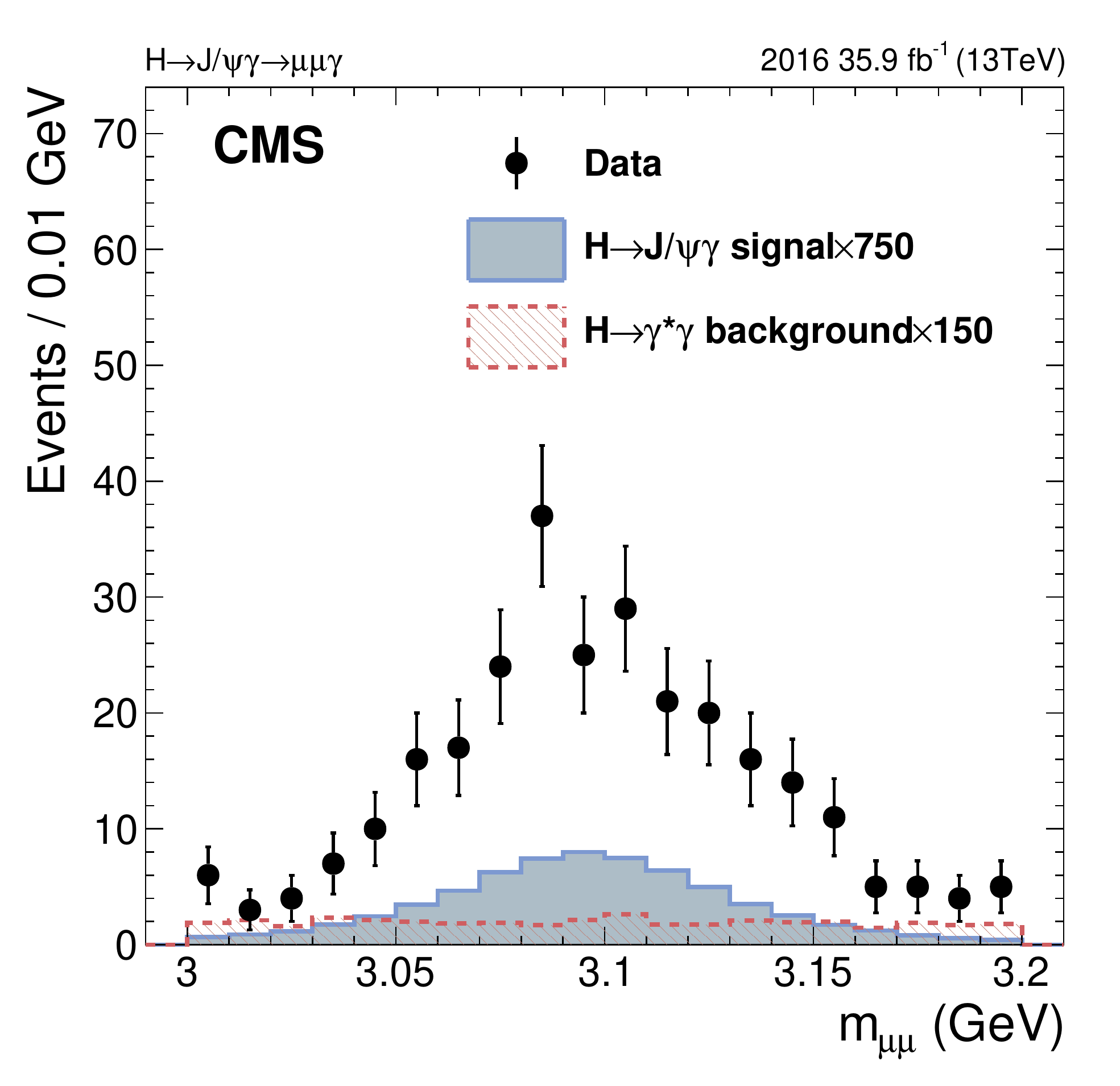}
    \caption{The $m_{\mu\mu}$ distributions in the $\cPZ$ (\cmsLeft) and Higgs (\cmsRight) boson searches. The number of events in the distributions from signal events is set to respective factors of 40 and 750 larger than the SM values for the predicted yields for $\cPZ$ and $\PH$ boson decays. The number of events in distributions in the resonant background samples is normalized to 5 and 150 multiples in  the expected yields. \label{fig:kineplot1}}
\end{figure}

\begin{figure}[!ht]
  \centering
    \includegraphics[width=0.45\textwidth]{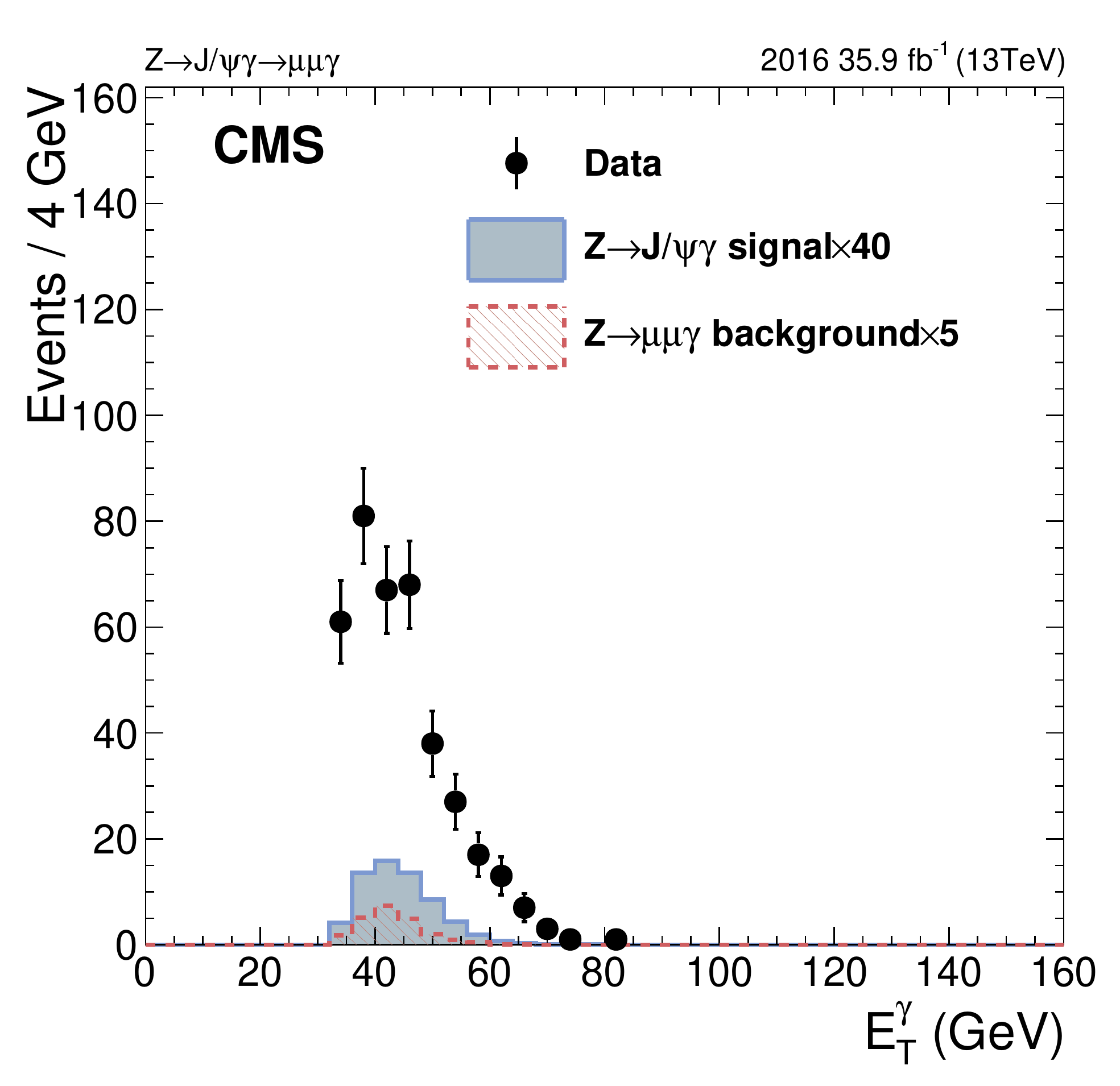}
    \includegraphics[width=0.45\textwidth]{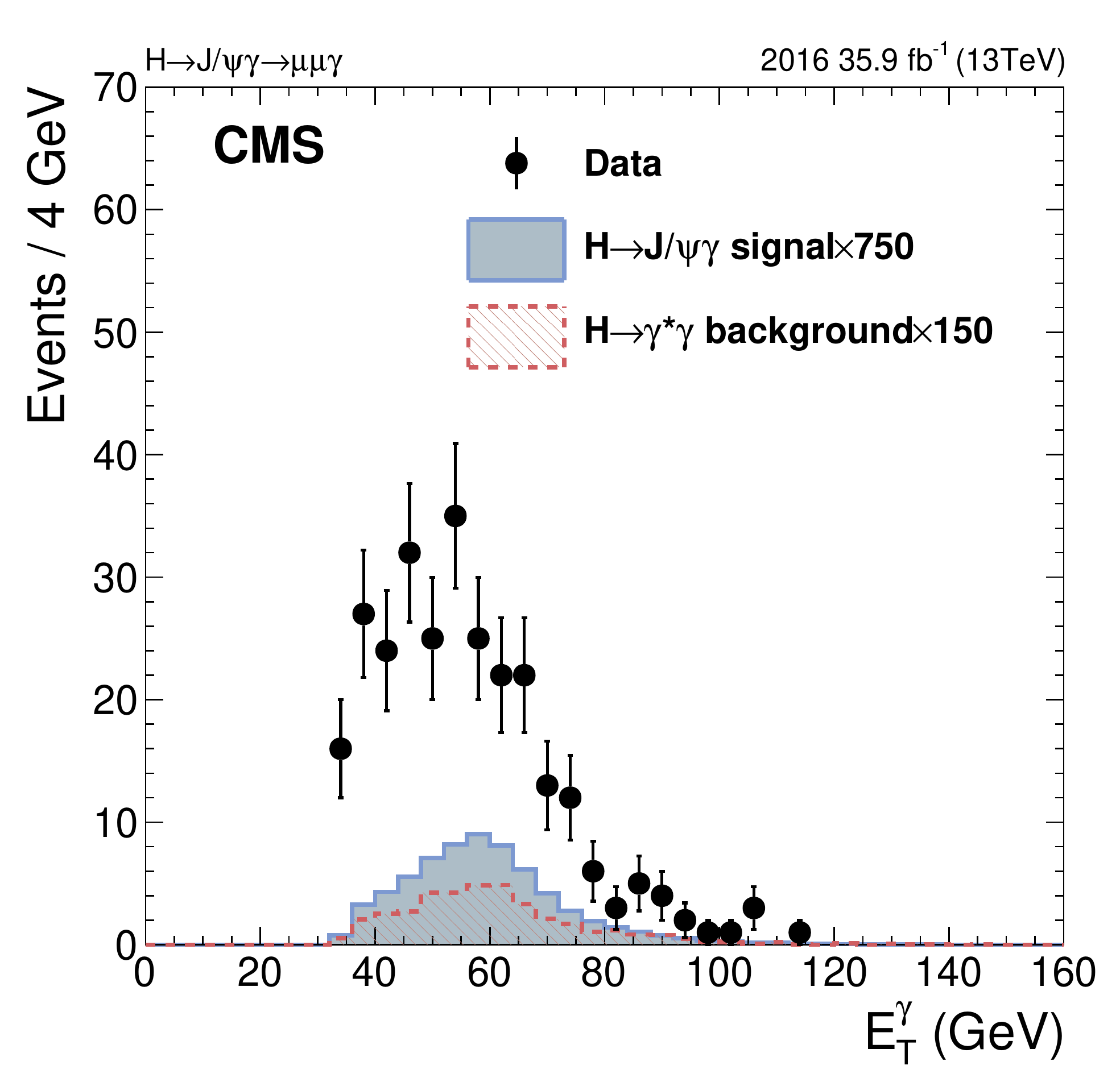}
    \caption{The photon $\et$ distributions in the $\cPZ$ (\cmsLeft) and Higgs (\cmsRight) boson searches. The number of events in the distributions from signal events is set to factors of 40 and 750 those of the SM predicted yields for the $\cPZ$ and $\PH$ boson decays, respectively. The number of events in distributions in the resonant background samples is normalized to respective factors of 5 and 150 larger than the expected yields. \label{fig:kineplot2}}
\end{figure}

\begin{figure}[!ht]
  \centering
    \includegraphics[width=0.45\textwidth]{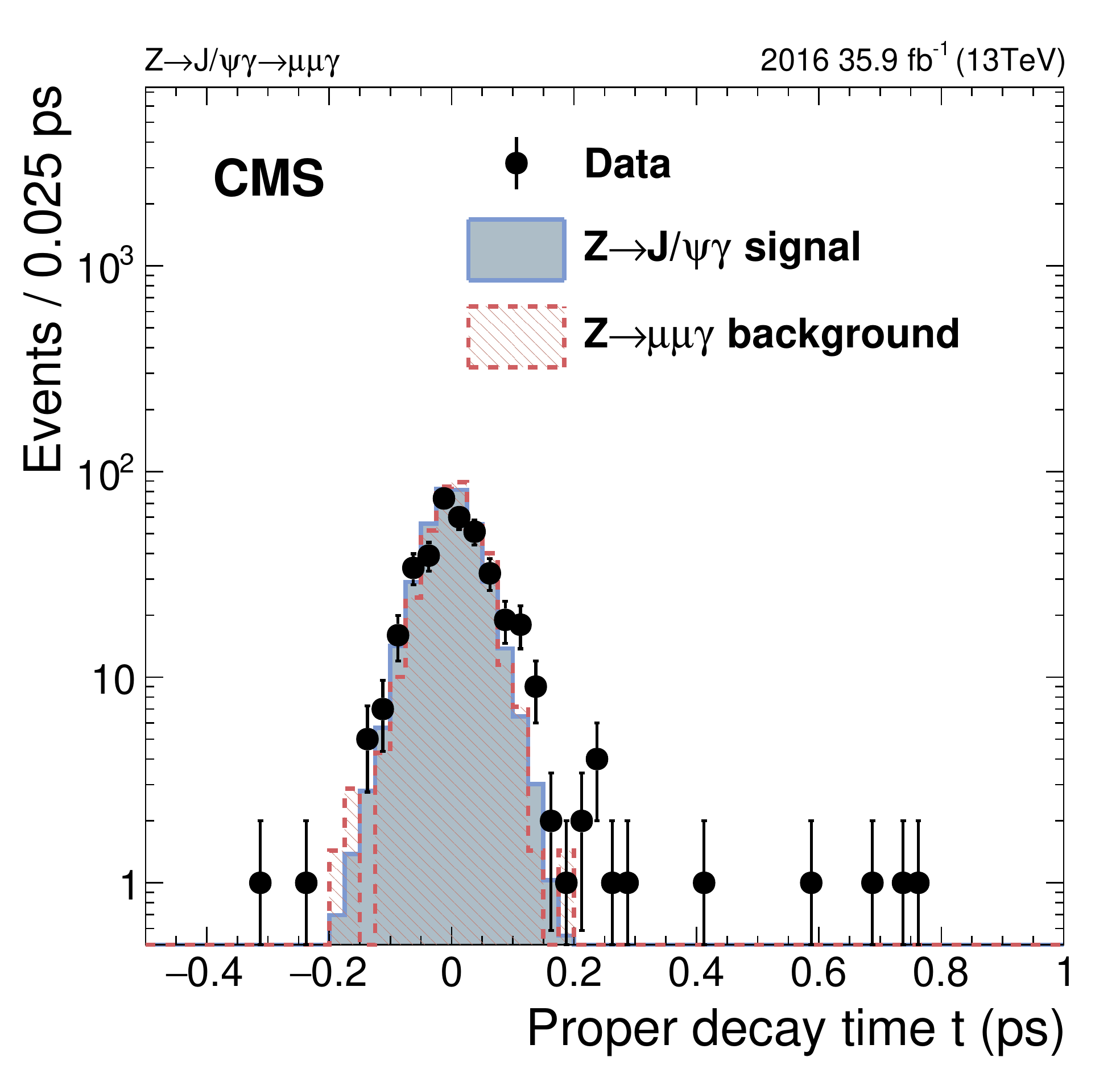}
    \includegraphics[width=0.45\textwidth]{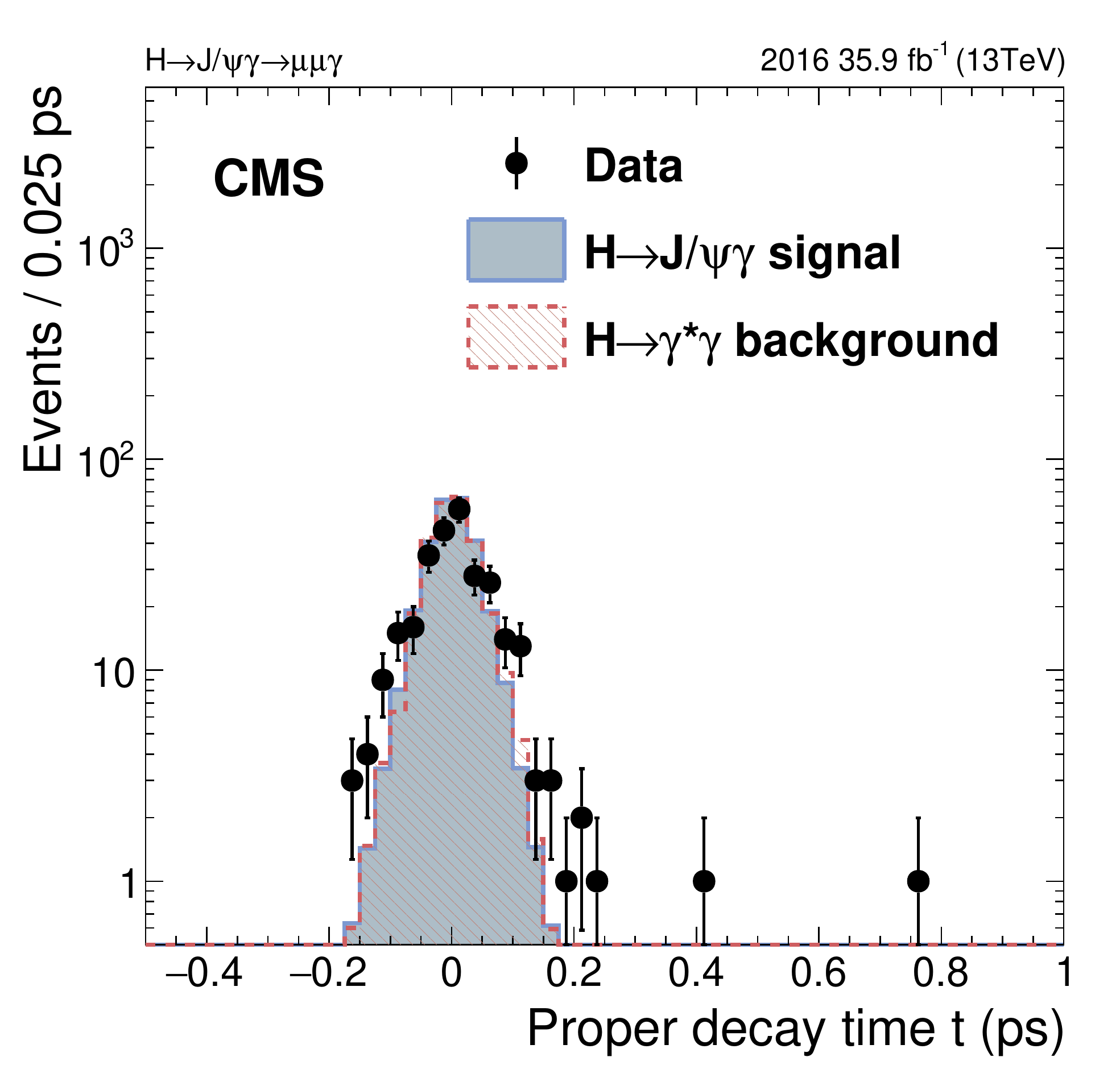}
    \caption{The proper decay time, t, distributions in the $\cPZ$ (\cmsLeft) and Higgs (\cmsRight) boson searches. Distributions in simulated events are normalized to the number of selected events in data. The distributions suggest that the $\JPsi$ candidates reconstructed in data, just as signal events, are produced promptly at the $\Pp\Pp$ interaction point, and not from displaced heavy-hadron decays. \label{fig:kineplot3}}
\end{figure}

Figures~\ref{fig:kineplot1} and \ref{fig:kineplot2} show the dimuon invariant mass and photon \et distributions for both $\cPZ$ and Higgs boson searches with events from all categories included. The number of events in the distributions from signal events is set to 40 (750) times the SM predicted yield for the $\cPZ$ ($\PH$) boson decay. The number of events in distributions in the resonant background samples is normalized to 5 (150) times the expected yield. The peak at the $\JPsi$ mass in data shows that real $\JPsi$ candidates are reconstructed and selected. These events come from inclusive quarkonium production; no simulation is available for this analysis so they cannot be included in the distributions. The background from $\cPZ\to\mu\mu\gamma$ events, for which a proper simulation exists, is much smaller than from inclusive quarkonium production, and it is scaled to make it visible. Figure~\ref{fig:kineplot3} shows the distribution of the proper decay time $t$, defined as $(m_{\mu\mu}/\pt^{\mu\mu}) L_{\mathrm{xy}}$, where $L_{\mathrm{xy}}$ is the distance between the primary event vertex and the common vertex of the muons in the transverse plane, for both $\cPZ$ and Higgs boson decays. These distributions are normalized to the number of selected events in data. The negative values come from the fact that $L_{\mathrm{xy}}$ is defined either to be positive or negative. The positive (negative) value indicates that the angle between the $L_{\mathrm{xy}}$ vector and the vector of $\pt^{\JPsi}$ is smaller (larger) than $\pi /2$. The distributions suggest that the $\JPsi$ candidates reconstructed in data, like the signal events, are produced promptly at the $\Pp\Pp$ interaction point, rather than coming from displaced heavy hadron decays.

\begin{table*}[!ht]
  \centering
    \topcaption{The number of observed $\cPZ$ or $\PH$ boson events, the expected signal yields, the expected nonresonant background with uncertainties estimated from the fit (described in Section~\ref{sec:model}), and the expected resonant background (see Section~\ref{sec:sample}) contribution in the ranges of 81 or 120 $< m_{\mu\mu\gamma} <$ 101 or 130\GeV, respectively, for the $\cPZ$ or $ \PH$ boson searches.\label{tab:Cutflow}}
    \cmsTable{
    \begin{tabular}{cccccccccc}
      \multicolumn{5}{c}{$\cPZ\to\JPsi\gamma$ ($81<m_{\mu\mu\gamma}<101\GeV$)} & \multicolumn{5}{c}{$\PH\to\JPsi\gamma$ ($120<m_{\mu\mu\gamma}<130\GeV$)}\\
    & Observed & & Nonresonant & Resonant & & Observed & & Nonresonant & Resonant \\
    Category & data & Signal & background & background & Category & data & Signal & background & background\\
    \hline
    EB high $\RNINE$ & 69 & 0.69 & 66.9$\pm 4.9$ & 2.1 & & & & \\
    EB low $\RNINE$ & 67 & 0.42 & 62.6$\pm 4.6$ & 1.2 & Inclusive & 56 & 0.076 & 51.0$\pm 3.4$ & 0.20\\
    EE & 47 & 0.30 & 43.0$\pm 4.0$ & 1.0 & & & & \\
    \end{tabular}
    }
\end{table*}

\section{Background and signal modeling}
\label{sec:model}
The subdominant, resonant backgrounds are estimated from the simulated samples, while the continuum background for each category for both the $\cPZ$ and Higgs boson decays is estimated and modeled using data by fitting a parametric function to the $m_{\mu\mu\gamma}$ distribution.  An unbinned maximum likelihood fit is performed over the range $70\ (100) < m_{\mu\mu\gamma} < 120\ (150)\GeV$ for the $\cPZ\ (\PH)\to\JPsi\gamma$ search.  The true form of the background $m_{\mu\mu\gamma}$ distribution is unknown and mismodeling of the background by the distribution obtained from the fit in data could lead to a bias in the analysis. The procedure used to study the bias introduced by the choice of function is described below.

{\tolerance=1200
Four families of functions are tested as potential parametrizations of the background: Bernstein polynomials, exponentials, power laws, and Laurent form polynomials. In the first step, one of the functions among the four families is chosen to fit the $m_{\mu\mu\gamma}$ distribution observed in data. Pseudo-events are randomly generated by using the resulting fit as a background model to simulate possible experiment results. Here, the order of the background function required to describe the data for each of the families is determined by increasing the number of parameters until an additional increase does not result in a significant improvement in the quality of the fit to the observed data. The improvement is quantified by the differences in the negative log-likelihood between fits with two consecutive orders of the same family of functions given the increment of the number of free parameters between two functions.
\par}

Signal events with signal strength $\mu_{\text{gen}}$ are introduced when generating the pseudo-events. The value $\mu_{\text{gen}}=1$ corresponds to injecting 1 times the signal yield expected from the SM on top of the sum of resonant and nonresonant background. A fit is made to the distribution using one of the functions in the four families combined with a signal model, where the normalization of the signal in this step is allowed to be negative. This procedure is repeated 5000 times and for each of the functions, and it is expected that ideally on average the signal strength predicted by the fit $\mu_{\text{fit}}$ will be equal to $\mu_{\text{gen}}$. The deviation of the mean fitted signal strength $\mu_{\text{fit}}$ from $\mu_{\text{gen}}$ in pseudo-events is used to quantify the potential bias. The criterion for the bias to be negligible is that the deviation must be at least five times smaller than the statistical uncertainty on $\mu_{\text{fit}}$. In other words, the distribution of the pull values, defined as $(\mu_{\text{fit}}-\mu_{\text{gen}})/\sigma_{\text{fit}}$, calculated from each pseudo-event should have a mean value of less than 0.2. This requirement implies and ensures that the uncertainty in the frequentist coverage, defined as the fraction of experiments where the true value is contained within the confidence interval, is negligible.

\begin{figure*}[htb]
  \centering
    \includegraphics[width=0.49\textwidth]{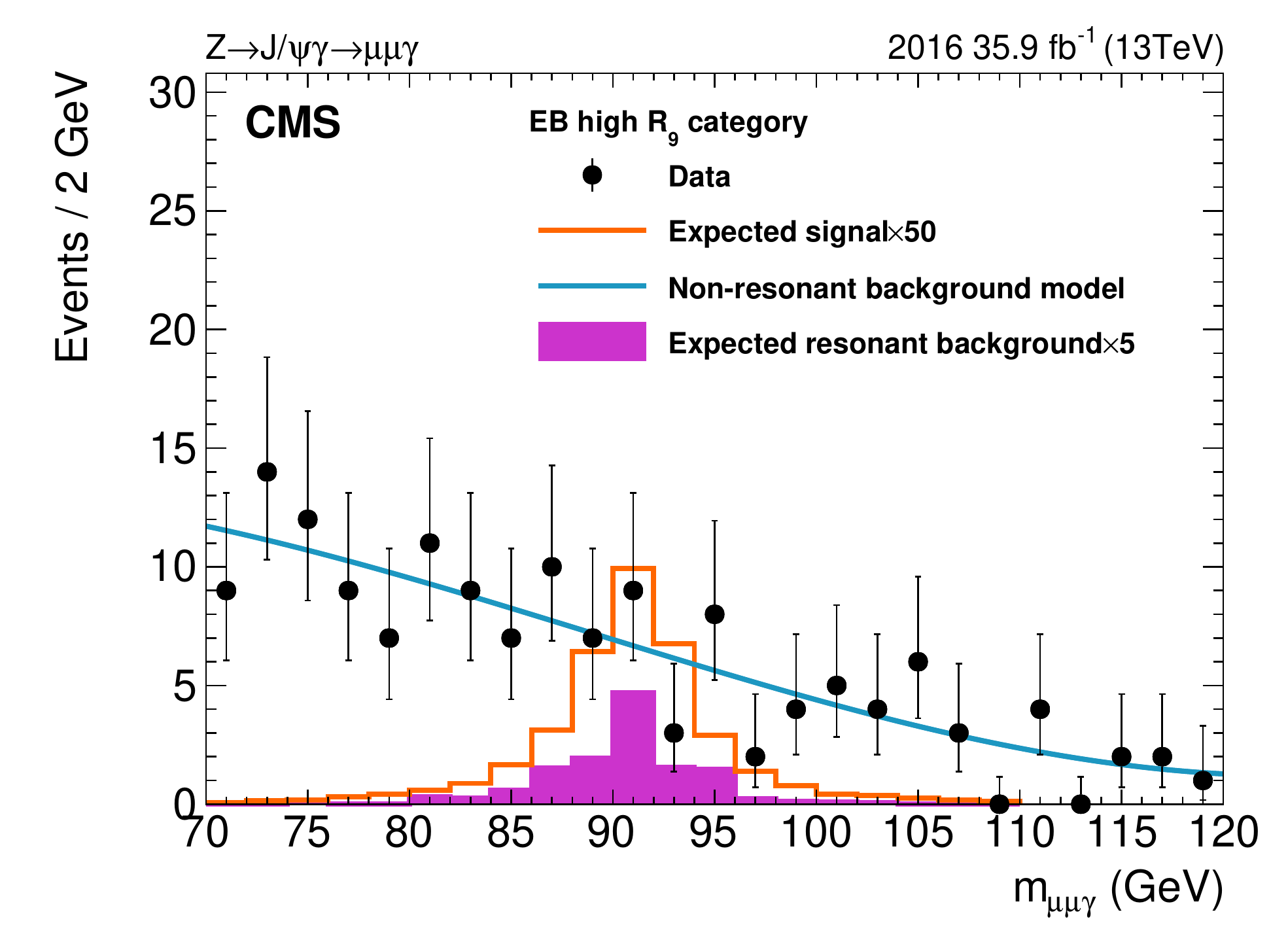} 
    \includegraphics[width=0.49\textwidth]{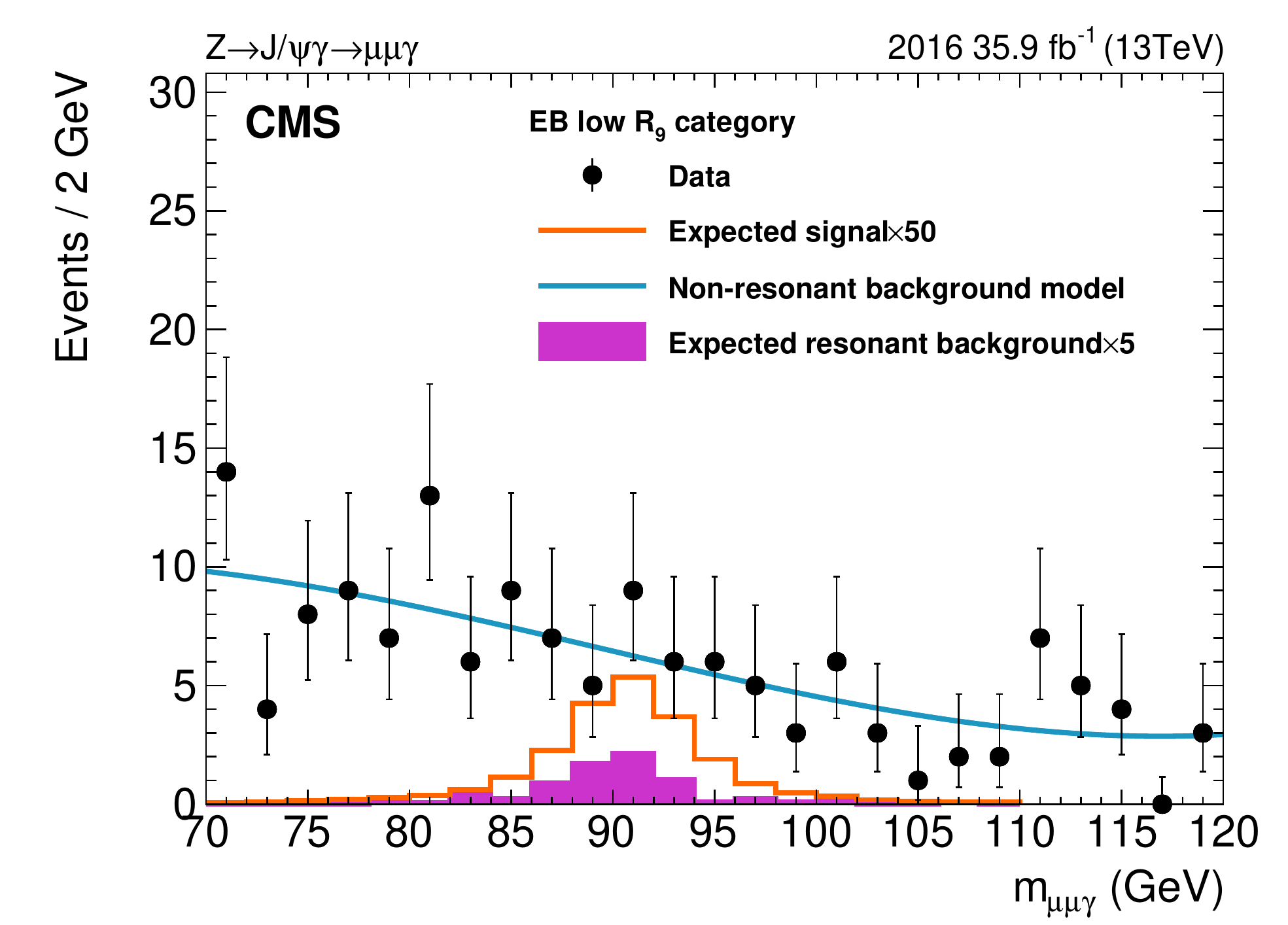}
    \includegraphics[width=0.49\textwidth]{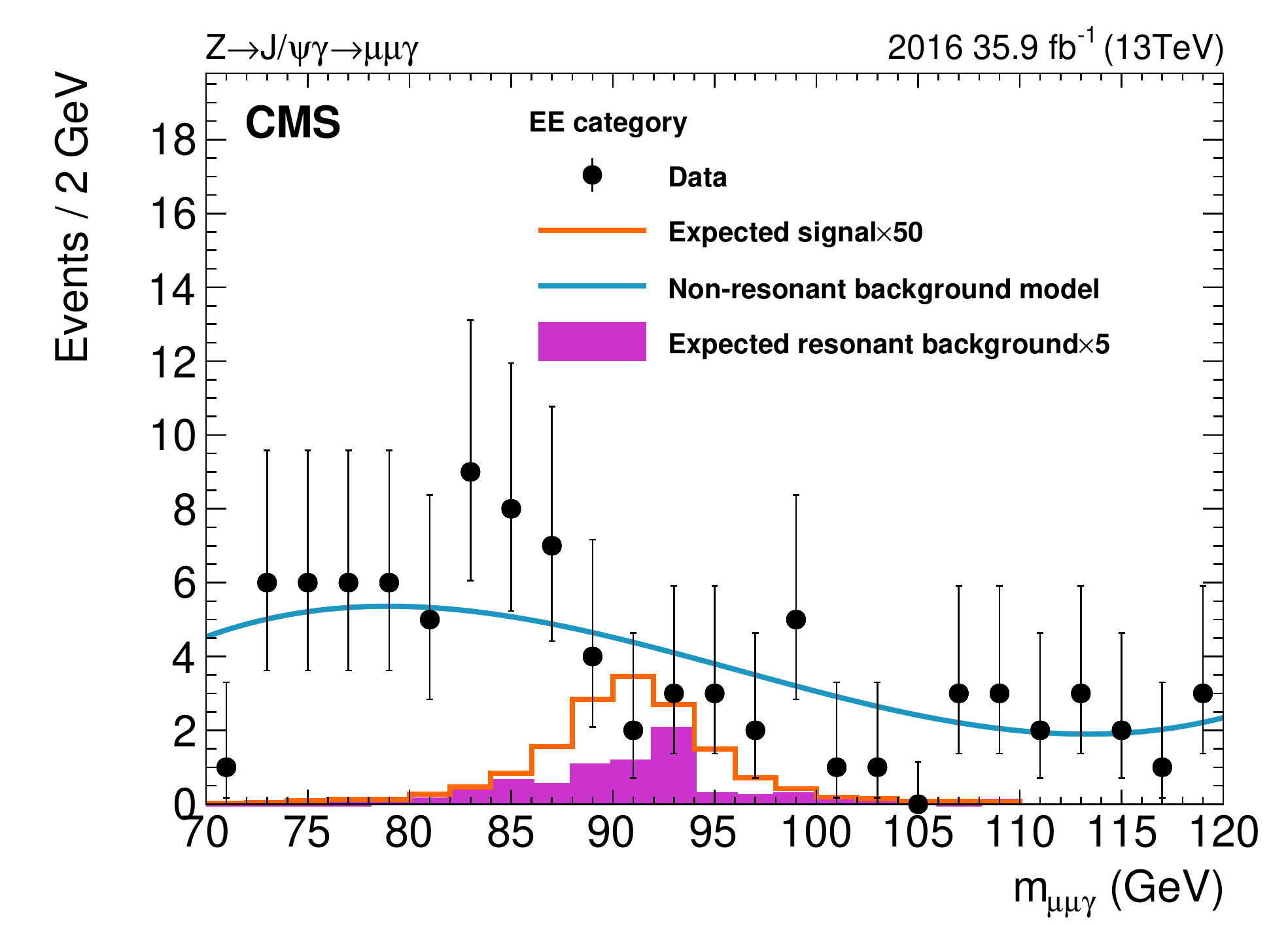}
    \includegraphics[width=0.49\textwidth]{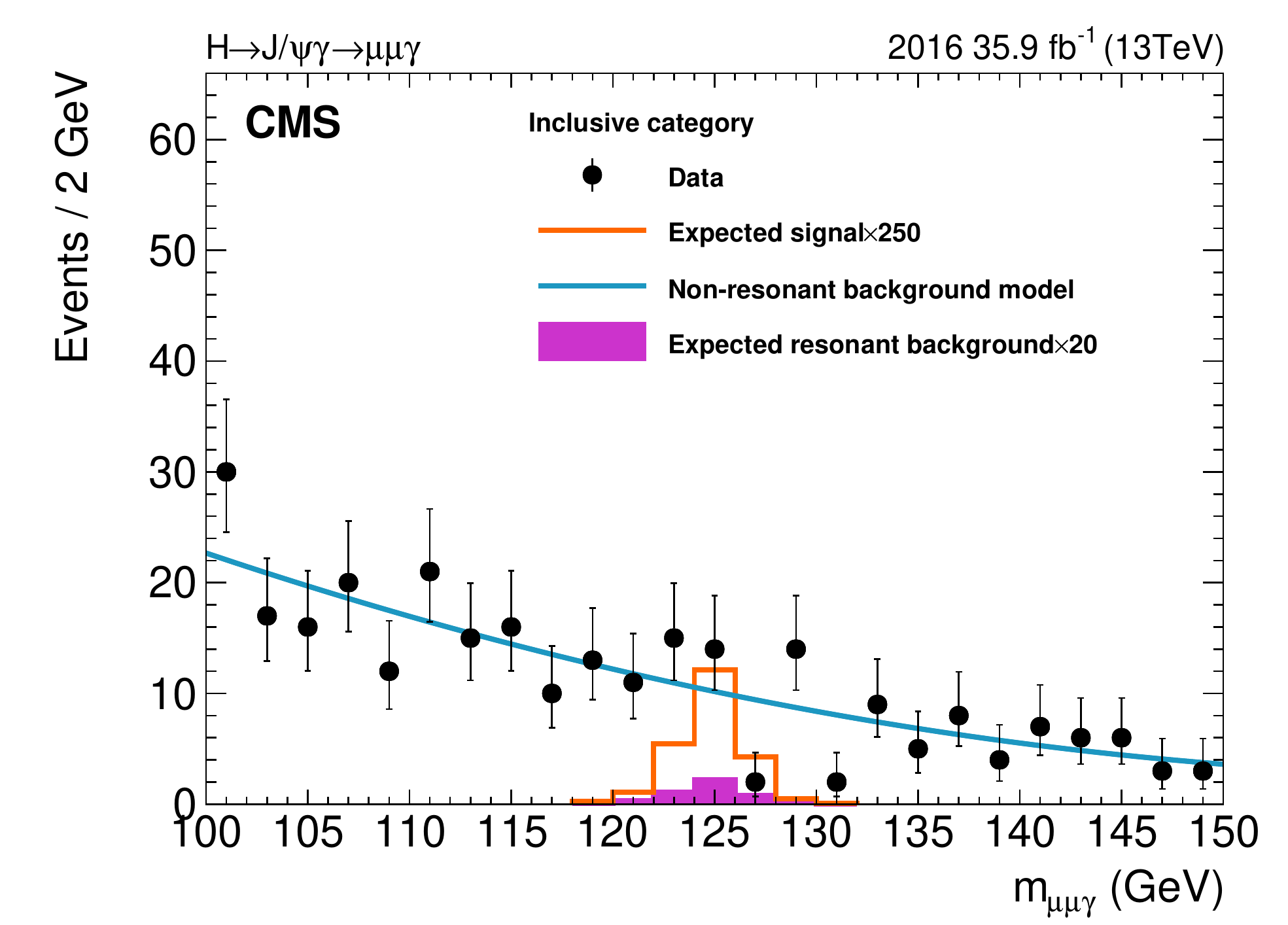}
    \caption[Multifit]{\label{fig:finalfit}
      Fits to nonresonant background using lowest-order unbiased functions to describe the three-body invariant mass $m_{\mu\mu\gamma}$ distributions observed in data for the $\cPZ\to\JPsi\gamma$ channel in the EB high $\RNINE$ category (top left), the EB low $\RNINE$ category (top right), the EE category (bottom left), as well as the $\PH\to\JPsi\gamma$ channel (bottom right).}
\end{figure*}

The polynomial background function satisfies the bias requirement. An order-three polynomial function is used for each category in the $\cPZ$ boson search, and an order-two polynomial function is used in the Higgs boson search. The $m_{\mu\mu\gamma}$ distribution and background model for each category is shown in Fig.~\ref{fig:finalfit}.

The signal model for each case is obtained from an unbinned maximum likelihood fit to the $m_{\mu\mu\gamma}$ distributions of the corresponding sample of simulated events. In the $\cPZ$ boson search, a double-sided Crystal Ball function~\cite{CB} is used. A Crystal Ball function plus a Gaussian with the same mean value is used in
the Higgs boson search.

\section{Results}
The distributions in $m_{\mu\mu\gamma}$ observed in the data are in agreement with the SM expectation of the background-only hypothesis. The results are used to derive upper limits on the branching fractions, $\mathcal{B}(\cPZ\to\JPsi\gamma)$ and $\mathcal{B}(\PH\to\JPsi\gamma)$. The exclusion limits are evaluated using the modified frequentist approach, $\CLs$, taking the profile likelihood as a test statistic~\cite{CLs_ref2,CLs_ref1,CLs_ref3,Cowan:2010js}. An unbinned evaluation of the likelihood is performed.

Systematic uncertainties in the expected number of signal events and in the signal model used in the fit come from the imperfect simulation of the detector and uncertainties in the theoretical prediction for the signal production. They are evaluated by varying contributing sources within their corresponding uncertainties and propagating the uncertainties to the signal yields or shapes in simulated signal samples. The sources of the uncertainties and their magnitudes are summarized in Table~\ref{tab:syst}. The uncertainties are classified into two types, one affecting the predicted signal yields and the other affecting the shapes of the signal models. The first type includes the uncertainties in the luminosity measurement~\cite{CMS-PAS-LUM-17-001}, the pileup modeling in the simulations, the corrections applied to the simulated events in order to compensate for differences in trigger, object reconstruction, and identification efficiencies, and the theoretical uncertainties. The theoretical uncertainties come from the effects of the PDF choice on the signal cross section~\cite{LHC-YR4,Ball:2014uwa,Butterworth:2015oua}, the lack of higher-order calculations for the cross-section~\cite{Martin2009,PhysRevD.82.074024,Alekhin:2011sk,Botje:2011sn,BALL2011296}, and the prediction of the decay branching fractions~\cite{Passarino}. The second type arises from the uncertainties in the momentum (energy) scale and resolution for muons (photons). These uncertainties are incorporated into the signal models by varying the momentum (energy) scale and resolution and introducing the effects on the mean and width of the Gaussian component of the signal models as shape nuisance parameters in the estimation of the limits.

The systematic uncertainties associated with the resonant background processes are evaluated with the methods used for the signal samples. The continuum background prediction is derived solely from data, so only statistical uncertainties are considered, which are translated into the uncertainties in each parameter of the fit function. The bias study mentioned in the previous section is performed to ensure that the bias from the choice of the background function is negligible. Hence, no additional systematic uncertainty is assigned to that background estimate.

The observed and median expected exclusion limits on the production cross sections and branching fractions at 95\% confidence level (\CL) for the $\cPZ$ and Higgs boson searches are summarized in Table~\ref{tab:limits_summary}. With the assumption that the $\JPsi$ meson is unpolarized, the observed upper limit on the branching fraction of $\cPZ\to\JPsi\gamma$ is $1.4\times 10^{-6}$, whereas the median expected upper limit is $1.6^{+0.7}_{-0.5}\times 10^{-6}$ with the 68\% \CL interval indicated by the subscript and superscript. The observed and median expected limits correspond to 15 and 18 times the SM prediction, respectively. Extreme polarization scenarios give rise to variations from -13.6 (-13.5)\%, for a fully longitudinally polarized $\JPsi$, to +8.6 (+8.2)\%, for a fully transversely polarized $\JPsi$ meson, in the observed (expected) branching fraction. The observed upper limit on the branching fraction of $\PH\to\JPsi\gamma$ is $7.6\times 10^{-4}$, and the median expected upper limit is $5.2^{+2.4}_{-1.6}\times 10^{-4}$. The observed and median expected limits correspond to 260 and 170 times the SM prediction. For the Higgs boson decay, the $\JPsi$ is assumed to be fully transversely polarized. The overall impact of systematic uncertainties in the final results is negligible.

The results from our $\PH\to\JPsi\gamma$ analysis are combined with the results from a similar search performed by the CMS Collaboration using $\Pp\Pp$ collision data at $\sqrt{s}=8\TeV$, corresponding to an integrated luminosity of 19.7\fbinv~\cite{Run1Paper_Dalitz}. The combination results in an upper limit corresponding to 220 (160) times the SM prediction.  The uncertainties are assumed either uncorrelated or correlated; the difference in the result is negligible.

\begin{table*}[!th]
  \centering
    \topcaption{Systematic uncertainties in both the searches for $\cPZ\to\JPsi\gamma$ and $\PH\to\JPsi\gamma$.  In the $\cPZ\to\JPsi\gamma$ search, the uncertainties are averaged over all categories. The numbers for uncertainties in the integrated luminosity, theoretical uncertainties, detector simulation and reconstruction correspond to the changes in the expected number of signal and resonant background events. The numbers for the uncertainties in the signal model correspond to the effect on the mean and width of the Gaussian component of the signal models resulting from the object momentum resolutions. \label{tab:syst}}
    \begin{tabular}{lcccc}
      & \multicolumn{2}{c}{$\cPZ\to\JPsi\gamma$ channel} & \multicolumn{2}{c}{$\PH\to\JPsi\gamma$ channel}\\
      & Signal & Resonant & Signal & Resonant \\
      Source &        & background &     & background \\
      \hline
      Integrated luminosity & \multicolumn{4}{c}{2.5\%}\\
      \multicolumn{5}{l}{Theoretical uncertainties} \\
      ~~~ Signal cross section (scale) & 3.5\% & 5.0\% &  \multicolumn{2}{c}{+4.6\% -6.7\%} \\
      ~~~ Signal cross section (PDF) & 1.7\% & 5.0\% & \multicolumn{2}{c}{3.2\%}\\
      ~~~ Branching fraction & \NA & 5.0\% & \NA & 6.0\%\\
      \multicolumn{5}{l}{Detector simulation, reconstruction}\\
      ~~~ Pileup weight & 0.8\% & 1.8\% & 0.7\% & 1.6\%\\
      ~~~ Trigger & 4.0\% & 4.0\% & 3.9\% & 4.0\%\\
      ~~~ Muon ident./Isolation & 3.0\% & 3.4\% & 2.0\% &2.5\% \\
      ~~~ Photon identification & 1.1\% & 1.1\% & 1.2\% &1.2\%\\
      ~~~ Electron veto & 1.1\% & 1.1\% & 1.0\% & 1.0\%\\
      \multicolumn{5}{l}{Signal model}\\
      ~~~ $m_{\mu\mu\gamma}$ scale  & 0.06\% & \NA & 0.1\% & \NA \\
      ~~~ $m_{\mu\mu\gamma}$ resolution  & 1.0\% & \NA & 4.8\% & \NA \\
    \end{tabular}
\end{table*}

\begin{table*}[!th]
  \centering
     \topcaption{Limits for $\cPZ$ and $\PH$ decays to $\JPsi\to\mu\mu$ final states. Shown in the second and third columns are the observed and expected limits for cross sections and branching fractions, with the upper and lower bounds in the expected 68\% \CL intervals shown, respectively, as superscripts and subscripts. The third column presents the $\cPZ$ decay branching fractions when the $\JPsi$ is assumed to be produced with $\lambda_{\theta}=+1$ or $-1$, in the helicity frame.  \label{tab:limits_summary}}
    \begin{tabular}{ ccccc }
    Channel & Polarization & $\sigma\ (\text{fb})$ at 95\% \CL & $\mathcal{B}(\cPZ\ (\PH)\to\JPsi\gamma)$ at 95\% \CL & $\frac{\mathcal{B}(\cPZ\ (\PH)\to\JPsi\gamma)}{\mathcal{B}_{\text{SM}}(\cPZ\ (\PH)\to\JPsi\gamma)}$\\
      \hline
      & Unpolarized & $4.6\ (5.3^{+2.3}_{-1.6})$ & $1.4\ (1.6^{+0.7}_{-0.5})\ten{-6}$ & 15 (18) \\
      $\cPZ\to\JPsi\gamma$ & Transverse & $5.0\ (5.9^{+2.5}_{-1.7})$ & $1.5\ (1.7^{+0.7}_{-0.5})\ten{-6}$ & 16 (19) \\
      & Longitudinal & $3.9\ (4.6^{+2.0}_{-1.4})$ & $1.2\ (1.4^{+0.6}_{-0.4})\ten{-6}$ & 13 (15) \\
      $\PH\to\JPsi\gamma$ & Transverse & $2.5\ (1.7^{+0.8}_{-0.5})$ & $7.6\ (5.2^{+2.4}_{-1.6})\ten{-4}$ & 260 (170) \\
    \end{tabular}
\end{table*}

\section{Summary}
A search is performed for decays of the standard model (SM) $\cPZ$ and Higgs bosons into a $\JPsi$ meson and a photon, with the $\JPsi$ meson subsequently decaying into $\PGmp\PGmm$. The data are from $\Pp\Pp$ collisions at $\sqrt{s}=13\TeV$, corresponding to an integrated luminosity of 35.9\fbinv. No excess is observed above the measured background. The observed and expected exclusion limits at 95\% confidence level (\CL) on the branching fraction of the $\cPZ$ boson decay in the unpolarized case are $\mathcal{B}(\cPZ\to\JPsi\gamma) < $ 1.4 and $1.6^{+0.7}_{-0.5}\ten{-6}$, corresponding to factors of 15 and 18 greater than the SM prediction. The 68\% \CL range in the confidence interval is shown as the subscript and superscript. Extreme polarization possibilities give rise to changes from -13.6 and -13.5\% for a longitudinally polarized $\JPsi$ meson, to +8.6 and +8.2\%, for a transversely polarized $\JPsi$ meson, in the respective observed and expected branching fractions. The 95\% \CL limit on the branching fraction of the Higgs boson are $\mathcal{B}(\PH\to\JPsi\gamma)<$ 7.6 and $5.2^{+2.4}_{-1.6}\ten{-4}$, corresponding to factors of 260 and 170 times the SM value. The results for the Higgs boson channel are combined with previous CMS data from proton-proton collisions at $\sqrt{s}=8\TeV$ to produce observed and expected upper limits on the branching fraction for the decay $\PH\to\JPsi\gamma$ of factors of 220 and 160 larger than the SM predictions.

\begin{acknowledgments}

We congratulate our colleagues in the CERN accelerator departments for the excellent performance of the LHC and thank the technical and administrative staffs at CERN and at other CMS institutes for their contributions to the success of the CMS effort. In addition, we gratefully acknowledge the computing centers and personnel of the Worldwide LHC Computing Grid for delivering so effectively the computing infrastructure essential to our analyses. Finally, we acknowledge the enduring support for the construction and operation of the LHC and the CMS detector provided by the following funding agencies: BMBWF and FWF (Austria); FNRS and FWO (Belgium); CNPq, CAPES, FAPERJ, FAPERGS, and FAPESP (Brazil); MES (Bulgaria); CERN; CAS, MoST, and NSFC (China); COLCIENCIAS (Colombia); MSES and CSF (Croatia); RPF (Cyprus); SENESCYT (Ecuador); MoER, ERC IUT, and ERDF (Estonia); Academy of Finland, MEC, and HIP (Finland); CEA and CNRS/IN2P3 (France); BMBF, DFG, and HGF (Germany); GSRT (Greece); NKFIA (Hungary); DAE and DST (India); IPM (Iran); SFI (Ireland); INFN (Italy); MSIP and NRF (Republic of Korea); MES (Latvia); LAS (Lithuania); MOE and UM (Malaysia); BUAP, CINVESTAV, CONACYT, LNS, SEP, and UASLP-FAI (Mexico); MOS (Montenegro); MBIE (New Zealand); PAEC (Pakistan); MSHE and NSC (Poland); FCT (Portugal); JINR (Dubna); MON, RosAtom, RAS, RFBR, and NRC KI (Russia); MESTD (Serbia); SEIDI, CPAN, PCTI, and FEDER (Spain); MOSTR (Sri Lanka); Swiss Funding Agencies (Switzerland); MST (Taipei); ThEPCenter, IPST, STAR, and NSTDA (Thailand); TUBITAK and TAEK (Turkey); NASU and SFFR (Ukraine); STFC (United Kingdom); DOE and NSF (USA).

\hyphenation{Rachada-pisek} Individuals have received support from the Marie-Curie program and the European Research Council and Horizon 2020 Grant, contract No. 675440 (European Union); the Leventis Foundation; the A. P. Sloan Foundation; the Alexander von Humboldt Foundation; the Belgian Federal Science Policy Office; the Fonds pour la Formation \`a la Recherche dans l'Industrie et dans l'Agriculture (FRIA-Belgium); the Agentschap voor Innovatie door Wetenschap en Technologie (IWT-Belgium); the F.R.S.-FNRS and FWO (Belgium) under the ``Excellence of Science - EOS" - be.h project n. 30820817; the Ministry of Education, Youth and Sports (MEYS) of the Czech Republic; the Lend\"ulet (``Momentum") program and the J\'anos Bolyai Research Scholarship of the Hungarian Academy of Sciences, the New National Excellence Program \'UNKP, the NKFIA research grants 123842, 123959, 124845, 124850 and 125105 (Hungary); the Council of Science and Industrial Research, India; the HOMING PLUS program of the Foundation for Polish Science, cofinanced from European Union, Regional Development Fund, the Mobility Plus program of the Ministry of Science and Higher Education, the National Science Center (Poland), contracts Harmonia 2014/14/M/ST2/00428, Opus 2014/13/B/ST2/02543, 2014/15/B/ST2/03998, and 2015/19/B/ST2/02861, Sonata-bis 2012/07/E/ST2/01406; the National Priorities Research Program by Qatar National Research Fund; the Programa Estatal de Fomento de la Investigaci{\'o}n Cient{\'i}fica y T{\'e}cnica de Excelencia Mar\'{\i}a de Maeztu, grant MDM-2015-0509 and the Programa Severo Ochoa del Principado de Asturias; the Thalis and Aristeia programs cofinanced by EU-ESF and the Greek NSRF; the Rachadapisek Sompot Fund for Postdoctoral Fellowship, Chulalongkorn University and the Chulalongkorn Academic into Its 2nd Century Project Advancement Project (Thailand); the Welch Foundation, contract C-1845; and the Weston Havens Foundation (USA).

\end{acknowledgments}

\bibliography{auto_generated}
\cleardoublepage \appendix\section{The CMS Collaboration \label{app:collab}}\begin{sloppypar}\hyphenpenalty=5000\widowpenalty=500\clubpenalty=5000\vskip\cmsinstskip
\textbf{Yerevan Physics Institute, Yerevan, Armenia}\\*[0pt]
A.M.~Sirunyan, A.~Tumasyan
\vskip\cmsinstskip
\textbf{Institut f\"{u}r Hochenergiephysik, Wien, Austria}\\*[0pt]
W.~Adam, F.~Ambrogi, E.~Asilar, T.~Bergauer, J.~Brandstetter, M.~Dragicevic, J.~Er\"{o}, A.~Escalante~Del~Valle, M.~Flechl, R.~Fr\"{u}hwirth\cmsAuthorMark{1}, V.M.~Ghete, J.~Hrubec, M.~Jeitler\cmsAuthorMark{1}, N.~Krammer, I.~Kr\"{a}tschmer, D.~Liko, T.~Madlener, I.~Mikulec, N.~Rad, H.~Rohringer, J.~Schieck\cmsAuthorMark{1}, R.~Sch\"{o}fbeck, M.~Spanring, D.~Spitzbart, A.~Taurok, W.~Waltenberger, J.~Wittmann, C.-E.~Wulz\cmsAuthorMark{1}, M.~Zarucki
% ...
\vskip\cmsinstskip
% ...
1:  Also at Vienna University of Technology, Vienna, Austria\\
% ...
\end{sloppypar}
\end{document}